\documentclass[journal,10pt, draftcls, onecolumn]{IEEEtran}
\IEEEoverridecommandlockouts
\usepackage{cite}
\usepackage{amsmath,amssymb,amsfonts}
\usepackage[caption=false,font=footnotesize]{subfig}
\usepackage{algorithm}
\usepackage{algorithmic}
\usepackage{graphicx}
\usepackage{textcomp}
\usepackage{xcolor}
\usepackage[binary-units]{siunitx}

\def\BibTeX{{\rm B\kern-.05em{\sc i\kern-.025em b}\kern-.08em
    T\kern-.1667em\lower.7ex\hbox{E}\kern-.125emX}}
\begin{document}

\title{A Simple Cooperative Diversity Method Based on Deep-Learning-Aided Relay Selection}

\author{Wei~Jiang,~\IEEEmembership{Senior~Member,~IEEE}
        and Hans~Dieter~Schotten,~\IEEEmembership{Member,~IEEE}

\thanks{\textit{Corresponding author: Wei Jiang (e-mail: wei.jiang@dfki.de)}}
\thanks{W. Jiang is with German Research Centre for Artificial Intelligence (DFKI), Kaiserslautern, Germany, and is also with the University of Kaiserslautern, Germany, (e-mail: wei.jiang@dfki.de).}
\thanks{H. D. Schotten is with German Research Centre for Artificial Intelligence (DFKI), Kaiserslautern, Germany, and is also with the University of Kaiserslautern, Germany, (e-mail: schotten@eit.uni-kl.de).}
}


\maketitle

\begin{abstract}
Opportunistic relay selection (ORS) has been recognized as a simple but efficient method for mobile nodes to achieve cooperative diversity in slow fading channels. However, the wrong selection of the best relay arising from outdated channel state information (CSI) in fast time-varying channels substantially degrades its performance.  With the proliferation of high-mobility applications and the adoption of higher frequency bands in 5G and beyond systems,  the problem of outdated CSI will become more serious. Therefore, the design of a novel cooperative method that is applicable to not only slow fading but also fast fading is  increasingly of importance.  To this end,  we develop and analyze a deep-learning-aided cooperative method coined predictive relay selection (PRS) in this article. It can remarkably improve the quality of CSI through fading channel prediction while retaining the simplicity of ORS by selecting a single opportunistic relay so as to avoid the complexity of multi-relay coordination and  synchronization. Information-theoretic analysis and numerical results in terms of outage probability and channel capacity reveal that PRS achieves full diversity gain in slow fading wireless environments and substantially outperforms the existing schemes in fast fading channels.
\end{abstract}


\section{Introduction}
\IEEEPARstart{I}{n} wireless communications \cite{Ref_Tse}, diversity is an important and essential technique, which can effectively combat the effect of multi-path channel fading by means of transmitting redundant signals over independent channels and then combining multiple faded copies at the receiver. Spatial diversity is particularly attractive as it can be easily combined with other forms of diversity and achieve higher diversity order by simply installing more antennas. Because of the constraint on power supply, hardware size, and cost, it is difficult for mobile terminals in cellular systems or wireless nodes in \emph{ad hoc} networks to exploit spatial diversity at sub-$6 \mathrm{GHz}$ carrier frequencies. Therefore, cooperative diversity  (cf. user cooperation diversity of \cite{Ref_usercoop}) has been proposed to break through this barrier. Exploiting the broadcast nature of radio signals in a relay channel \cite{Ref_RelayChannel}, cooperating terminals share their distributed antennas to form `a virtual array'. In such a cooperative network, when a node sends a signal, its neighboring nodes could act as relays to \emph{decode-and-forward} (DF) or \emph{amplify-and-forward} (AF) this signal. By combining multiple copied versions of the original signal at the destination, the network achieves cooperative diversity that is equivalent to spatial diversity gained from co-located multi-antenna systems \cite{Ref_massiveMIMO}.

To  achieve cooperative diversity, a cooperation strategy is required to rule which nodes should participate in relaying and how to collaboratively retransmit? The repetition-based cooperative strategies presented in \cite{Ref_CoopDiversity} simply repeat the signal on orthogonal channels to realize full diversity, but this gain comes with a price of substantial loss on spectral efficiency. To avoid this penalty, a method called distributed beamforming has been discussed in \cite{Ref_usercoop, Ref_DistributedBF}. Assuming \emph{a priori} knowledge of forward channels, the source and relays could simultaneously  transmit signals for a coherent combination at the receiver. Beamforming is vulnerable to phase noise, whereas radio-frequency-chain calibration among distributed antennas (relays) to align phase distortion is difficult to implement.  In \cite{Ref_DSTC}, an approach called distributed space-time coding\,(DSTC) has been proposed. Although full diversity on the order of the number of relays can be achieved, designing such a code is still an open issue since the number of distributed antennas is unknown and time-varying. Additionally, multiple timing offset (MTO) \cite{Ref_SYN01} and multiple carrier frequency offset (MCFO) \cite{Ref_SYN02} among spatially-distributed relays make the aforementioned \emph{multi-relay} transmission too complicated for practical systems.

Inspired by the benefit of \textit{selection diversity} from multi-user selection \cite{Ref_multiuserSelectionP} and antenna selection \cite{Ref_Feedbackdelay}, relay selection was proposed to simplify the implementation of cooperative networks. In \cite{Ref_zhao2005practical}, a location-based approach that selects the best relay based on ideas from geographical random forwarding \cite{Ref_zorzi2003geographic} was presented. Assuming that each node knows its own position, as well as that of the destination, the node
closest to the destination serves as the relay. Such schemes are more appropriate for static networks but less appropriate for mobile networks because the estimation of positions or distances among all nodes is not a trivial task. 
In contrast, another \emph{single-relay} approach referred to as opportunistic relay selection (ORS) \cite{Ref_OPR}  that requires no topology information  was proposed.     Using local channel measurements, this approach opportunistically selects a single relay with the best channel condition (in accordance to a given selection criterion \cite{Ref_OPR2}). From the viewpoint of multiplexing-diversity trade-off, ORS has no performance loss compared to more complex protocols such as DSTC. Most importantly, it substantially lowers the complexity of implementation by avoiding synchronization among multiple transmitting relays  while the requirement of space-time codes is completely eliminated. It was recognized as a simple but efficient way to achieve cooperative diversity in \textit{slow} fading channels. 
In \textit{fast} fading wireless environment, however, the measured channel state information (CSI) for relay selection may differ from the actual channel quality at the instant of signal relaying due to processing and feedback delay. The outdated CSI causes wrong relay selection, which drastically deteriorates the performance of ORS, as extensively verified in \cite{Ref_Vicario, Ref_Seyfi, Ref_Torabi_Capacity01, Ref_Torabi, Ref_PartialRelaySelection}. With the proliferation of high-mobility applications  and the adoption of higher frequency bands in 5G and beyond systems,  the problem of outdated CSI will become more serious. According to the Doppler effect in signal propagation \cite{Ref_Tse}, transmitting signals at higher frequency (such as millimeter wave and Terahertz communications) or moving at a higher speed (e.g., vehicular communications, high-speed trains, and unmanned aerial vehicles) will increase the frequency shift, leading to a faster time-varying channel.  Hence, the design of a simple cooperative method that can be also applicable to fast fading channels is  increasingly of significance for next-generation wireless communications.

To the best knowledge of the authors, a few proposals for cooperative diversity in the presence of outdated CSI have been reported in the literature.  Generalized selection combining and its enhanced version \cite{Ref_Xiao, Ref_Chen, Ref_Jiang_GSC}, which select $N$ relays with good channel quality to retransmit in an orthogonal manner, exhibit robustness in the presence of outdated CSI, whereas its loss of spectral efficiency to $1/N$ is not acceptable. The authors of \cite{Ref_Li_RelaySelection} proposed a method utilizing the knowledge of channel statistics.  It gets only marginal performance improvement, but the complexity obviously grows. In \cite{Ref_MineTRANS, Ref_Jiang_ICC}, one author of this article designed a scheme called opportunistic space-time coding (OSTC) that combines the benefits of both \textit{opportunistic} relaying and distributed \textit{space-time coding}.  A fixed number of $N$ relays are opportunistically selected and $N$-dimensional orthogonal space-time block coding is employed on these relays. It can improve the performance of cooperative networks over fast fading channels while avoiding the loss of spectral efficiency. However,  its performance gap away from the full diversity achieved by using perfect CSI is still large, motivating our works presented in this article.

Channel prediction \cite{Ref_ChannelPredictionReview, Ref_myIEEEAccess, Ref_AR2}, which can improve the timeliness of CSI without spending radio resources, is promising to combat outdated CSI.  It earns a prediction horizon that can be used to counteract induced delay. Modeling a wireless channel into a set of propagation parameters, two statistical predictive approaches  - auto-regressive \cite{Ref_AR3} and parametric model \cite{Ref_Parametric02} -  have been proposed. But these models are fossilized, leaving a gap from real channels, and - in addition - the parameter estimation relying on complex algorithms such as MUSIC and ESPRIT \cite{Ref_MUSIC} is tedious, harmed its applicability in practical systems  \cite{Ref_Wei_ICC2019}. In 2016, when AlphaGo \cite{Ref_AlphaGo},  a deep learning (DL) computer program, achieved a historic victory versus a human champion, the passion of exploring Artificial Intelligence (AI) in almost every scientific and engineering branches was ignited  \cite{Ref_PIMRC}. As an important AI technique,  recurrent neural networks show strong capability on time-series prediction \cite{Ref_TimeSeriesPrediction} and are applied to provide a data-driven alternative to efficiently implement wireless channel prediction \cite{Ref_myOJCOMS,  Ref_myBook_NN, Ref_JiangFDpredict}.

Taking advantage of new degree of freedom opened by channel prediction,   we develop and analyze a novel cooperative diversity method coined predictive relay selection (PRS) in this article. Its key idea is to apply a DL-based channel predictor to improve the quality of CSI so as to lower the probability of wrong relay selection. To this end, a deep recurrent network that specifically adapts to the characteristics of CSI data  is elaborately built.  To avoid MTO and MCFO in multi-relay transmission, only a single relay is opportunistically selected in terms of  predicted CSI.  Frame structures supporting for either distributed or centralized PRS are designed accordingly.   Information theoretic analysis is conducted by deriving closed-form expressions for outage probability and channel capacity, which are corroborated by simulation results. Moreover, its computational complexity, robustness, and scalability are investigated.
The contributions and organization of this article are listed as follows:
\begin{enumerate}
    \item Section II models a half-duplex dual-hop cooperative network using either AF or DF relays, and reviews the existing schemes including ORS and OSTC.
    \item Section III  provides the principle of deep recurrent neural networks and the methodology to build channel predictors. The statistics of predicted CSI and the computational complexity for predictors are analyzed.
    \item Section IV presents the proposed scheme and the design of two frame structures for distributed and centralized PRS, respectively.
    \item In Section V and VI, information-theoretic analyses for the proposed scheme in both AF and DF relaying are conducted through deriving  closed-form expressions of outage probability and channel capacity.
    \item In Section VII, the acquisition of CSI dataset  and the selection of hyper-parameters for high-accuracy prediction are clarified. Performance evaluation is carried out through Monte-Carlo simulations to corroborate the theoretical analyses. Moreover, we study its robustness against additive noise, synchronization error, mobility, and fading statistics, scalability in terms of the number of relays, and computational complexity in comparison with the capability of commercial off-the-shelf (COTS) computing hardware.
    \item Finally, Section VIII concludes this article.
\end{enumerate}

\textbf{Notations}:: Throughout this article, bold lower-case and upper-case letters denote vectors and matrices, respectively. For their operation,  $(\cdot)^*$, $(\cdot)^T$,  and $(\cdot)^H$ notate the conjugate, transpose, and Hermitian transpose, respectively, $\| \cdot\|$ expresses the Frobenius norm, and $\otimes$ marks the Hadamard (element-wise) product. $\mathbb{E}$ denotes the statistical expectation, $\mathbb{P}$ is the notation of mathematical probability,  $\Re$ and $\Im$ take the real and imaginary units of a complex quantity. $\mathbf{h}$, $\mathbf{\hat{h}}$, and $\mathbf{\check{h}}$ represent  the actual, outdated, and predicted CSI, respectively.

\section{System Model}
Following the working assumption for the majority of prior research works in \cite{Ref_DSTC, Ref_DistributedBF, Ref_SYN01, Ref_SYN02, Ref_OPR, Ref_OPR2, Ref_Vicario, Ref_Seyfi, Ref_Torabi_Capacity01, Ref_Torabi,  Ref_PartialRelaySelection}, we consider a dual-hop cooperative network where a single source node $s$ communicates with a single destination node $d$ with the aid of $K$ relays, neglecting the direct link at the destination for simplifying the analysis\footnote{With a direct link, the overall signal-to-noise ratio (SNR) is $\gamma_{tot}=\gamma_{s,d}+\gamma_{k,d}$ under maximal-ratio combining at the receiver, where $\gamma_{s,d}$ and $\gamma_{k,d}$ are the SNRs of the direct and relay link. Its achievable diversity order is $K+1$ and $2$ with the prefect and outdated CSI, respectively, compared to $K$ and $1$ in the case of no direct link. Neglecting the direct link does not affect the performance impact of outdated CSI on the relay selection, as illustrated by the results in the simulation section.}. Each node is equipped with a single antenna that is used for both signal transmission and reception over a narrow-band channel. Although the proposed scheme is applicable for any kind of wireless channel statistics, without loss of generality, we adopt Rayleigh fading to analyze performance for simplicity. Thus, the channel realization is a zero-mean circularly-symmetric complex Gaussian random variable with variance  $\sigma_h^2$, i.e., $h {\sim} \mathcal{CN}(0, \sigma_h^2)$. The received signal in an arbitrary link $A{\rightarrow}B$ is modeled as $y_{B} = h_{A,B}x_{A} + z_{B}$, where $x_A\in \mathcal{C}$ is the transmitted symbol from node $A$ with average power $P_A=\mathbb{E}[|x_A|^2]$,  $h_{A,B}$ represents the fading coefficient of the channel from $A$ to $B$, and $z_{B}$ stands for additive white Gaussian noise with zero-mean and variance $\sigma^2_n$, i.e., $z {\sim} \mathcal{CN}(0,\sigma^2_n)$. The instantaneous signal-to-noise ratio (SNR)  is denoted by $\gamma_{A,B}{=}|h_{A,B}|^2 P_A /\sigma_n^2$ and the average SNR $\bar{\gamma}_{A,B}{=}\mathbb{E}[\gamma_{A,B}]{=}\sigma_h^2 P_A/\sigma_n^2$. Node $A$ can be the source $A=s$ or a relay $A=k$, $k{\in}\{1,...,K\}$, corresponding to  $B=k$ or $B=d$. It is noted that relay selection depends on instantaneous channel realizations or equivalently on received instantaneous SNRs, which are interchangeably used in the context of relay selection hereinafter.

From a practical point of view, there exists a delay between the time of relay selection and the instant of using the selected relay to transmit. The actual CSI $h$ may differ from its outdated version $\hat{h}$ that is applied for selecting relays. To quantify the quality of CSI, the correlation coefficient between $h$ and $\hat{h}$ is introduced, i.e.,
\begin{equation}
\label{Eqn_CorCoeff}
\rho_o=\frac{\mathbb{E}[h\hat{h}^*]}{\sqrt{\mathbb{E}[|h|^2] \mathbb{E}[|\hat{h}|^2]}}.
\end{equation}
According to \cite{Ref_Feedbackdelay}, we have
$\hat{h}=\sigma_{\hat{h}} \left ( \frac{\rho_o}{\sigma_h} h + \varepsilon \sqrt{1-\rho_o^2}  \right )$, where $\varepsilon$ is a random variable with standard normal distribution $\varepsilon \sim \mathcal{CN}(0,1)$ and $\sigma_{\hat{h}}^2$ is the variance of $\hat{h}$.
With the classical Doppler spectrum of the Jakes model, it takes the value
\begin{equation}
\label{Eqn_Jakes}
\rho_o=J_0(2\pi f_d \tau),
\end{equation}
where $f_d$ is the maximal Doppler frequency, $\tau$ stands for the delay between the outdated and actual CSI, and $J_0(\cdot)$ denotes the $zeroth$ order  Bessel  function  of the first kind.

\begin{figure}[!htbp]
\centering
\includegraphics[width=0.48\textwidth]{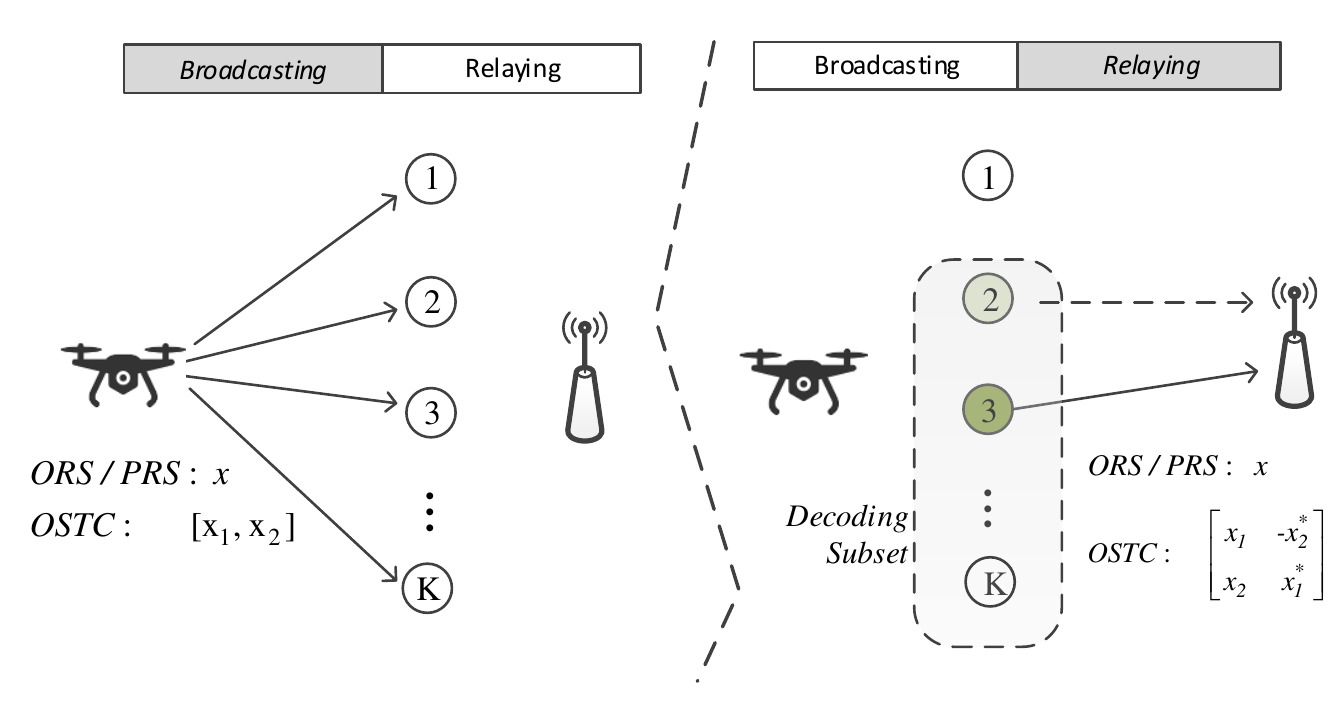}
\caption{Schematic diagram of a cooperative network with different DF relaying strategies: ORS, PRS, and OSTC. In the $1^{st}$ phase, the source broadcasts a signal, while the relays that successfully decode this signal form a $\mathcal{DS}$. In the $2^{nd}$ phase, the selected node(s)  from the $\mathcal{DS}$ forwards the regenerated signal.  Examples of deployment scenarios for such a cooperative network include: a flying drone suffering from sparse signal coverage maintains its connectivity to a network via a group of ground terminals; a platoon of moving vehicles optimize their mutual communications via relaying; a set of Internet-of-Things (IoT) devices collaboratively improve the reliability to access an edge server; and, a few neighboring user terminals in a cell, especially at cell edge, cooperatively boost their performance in uplink.     }
\label{Figure_RelaySele}
\end{figure}

\subsection{Decode-and-Forward}
Due to severe signal attenuation, a single-antenna relay should operate in half-duplex mode to prevent from harmful self-interference between the transmitter and receiver. Without loss of generality, orthogonal transmission between the source and relays using time-division multiplexing is used for analysis throughout the sequel (while frequency-division multiplexing can also be equivalently applied). Therefore, its signal transmission is organized in two phases:  the source broadcasts a signal in the source-to-relay (denoted by $\mathbb{SR}$ hereinafter)  link, and then the relays retransmit this signal in the relay-to-destination ($\mathbb{RD}$) link.
In the first phase, as shown in \figurename \ref{Figure_RelaySele}, the source (e.g., the drone in the figure) sends a symbol $x$ and those relays which  overhear and correctly decode this signal  form a \emph{decoding subset} ($\mathcal{DS}$) of the $\mathbb{SR}$ link
\begin{IEEEeqnarray}{ll}
\label{Eqn_DS} \nonumber
\mathcal{DS}  &\triangleq \left \{k \left| \frac{1}{2} \log_2(1+\gamma_{s,k}) \geqslant R  \right\}\right. \\
                 &=  \{k \left| \gamma_{s,k} \geqslant \gamma_{o}  \}\right. ,
\end{IEEEeqnarray}
where $R$ is an end-to-end target rate for the dual-hop relaying, corresponding to a threshold SNR  $\gamma_{o}=2^{2R}{-}1$. Note that the required data rate for either hop is doubled to $2R$ due to the adoption of half-duplex  transmission.

The best relay (denoted by $\dot{k}$) in the conventional ORS  is opportunistically selected from $\mathcal{DS}$ in terms of $\dot{k}=\arg \max_{k\in \mathcal{DS}} {\hat{\gamma}_{k,d}}$, where $\hat{\gamma}_{k,d}$ is the SNR of the $\mathbb{RD}$ link \emph{at the instant of relay selection}, which is an outdated version of the actual SNR $\gamma_{k,d}$ during signal transmission. In contrast, the proposed PRS scheme replaces  outdated CSI with  predicted CSI $\check{h}$, and determines $\dot{k}$ in terms of $\dot{k}=\arg \max_{k\in \mathcal{DS}} {\check{\gamma}_{k,d}}$, where $\check{\gamma}_{k,d}=|\check{h}_{k,d}|^2P_k/\sigma_n^2$. In addition to the best relay, the OSTC scheme \cite{Ref_MineTRANS} needs another relay with the second strongest SNR, i.e., $\ddot{k}=\arg \max_{k\in \mathcal{DS}-\{\dot{k}\}} {\hat{\gamma}_{k,d}}$.
In the first phase, the source broadcasts a pair of symbols $(x_1,x_2)$ over two consecutive symbol periods. The regenerated symbols are encoded by means of the Alamouti scheme,  which is the unique space-time code achieving both full rate and full diversity, at the pair of selected relays. In the second phase, a relay transmits $(x_1,-x_2^*)$ while another transmits $(x_2,x_1^*)$ simultaneously at the same frequency over two symbol periods.

\subsection{Amplify-and-Forward}
Compared to DF, the main difference of AF is that the best relay does not detect the received signal, while only amplifying it.
In the first phase, the source broadcasts $x$, and thus the received signal at the $k^{th}$ relay is $
y_{k}  =  h_{s,k}x + z_{k}$. Relay $k$  normalizes  $y_{k}$ to form a retransmitted signal:
\begin{equation}
  x_{k} = \frac{\sqrt{P_k} y_{k}}{ \sqrt{\mathbb{E}[ |y_{k}|^2 ]}}  =\frac{\sqrt{P_k} (h_{s,k } x + z_{k} ) }{ \sqrt{P_s |h_{s,k }|^2   + \sigma_n^2 }  },
\end{equation}
where $P_s{=}\mathbb{E}[|x|^2]$ is the average transmit power of the source and $P_k{=}\mathbb{E}[|x|_k^2]$ is the average power for retransmission.
The receiver at the destination gets
\begin{equation}
\label{forular_AFmodel}
  y_{d} = h_{k,d} x_{k} + z_{d} =  \frac{ \sqrt{P_k} h_{k,d}(h_{s,k } x + z_{k}) }{ \sqrt{ P_s |h_{s,k }|^2  + \sigma_n^2 }  }  + z_{d}.
\end{equation}
Thus, the received SNR for this end-to-end ($\mathbb{EE}$) link is
\begin{equation}
\label{Eqn_xx1}
\gamma_{skd} = \frac{\gamma_{s,k}\gamma_{k,d}}{\gamma_{s,k}+\gamma_{k,d}+1},\:\:\:k\in\{1,...,K\}.
\end{equation}
For the sake of mathematical tractability, as recommended in \cite{Ref_Torabi}, a tight upper bound  is used to approximate (\ref{Eqn_xx1}), that is
\begin{equation}
\label{Eqn_minlink}
\gamma_{skd} \leqslant \gamma_{k} = \min \{ \gamma_{s,k}, \gamma_{k,d} \}, \:\: k\in\{1,...,K\}.
\end{equation}
As explained previously, the instantaneous SNR used for relay selection is an outdated version of (\ref{Eqn_minlink}), i.e., $\hat{\gamma}_{k} = \min \{ \hat{\gamma}_{s,k}, \hat{\gamma}_{k,d} \}$. The ORS scheme \cite{Ref_Torabi_Capacity01} opportunistically selects the best path out of $K$ possible $\mathbb{EE}$ links, we have $\dot{k}=\arg \max_{k\in\{1,...,K\}} \left\{\hat{\gamma}_{k}\right\}$.
In contrast, the proposed scheme chooses the best relay as $\dot{k}=\arg \max_{k\in\{1,...,K\}}\left\{ {\min(\check{\gamma}_{s,k}}, \check{\gamma}_{k,d})\right\}$, where $\check{\gamma}_{s,k}$ and $\check{\gamma}_{s,k}$ are predicted CSI for the $\mathbb{SR}$ and $\mathbb{RD}$ link, respectively.

\section{Deep learning-based Channel Prediction}
This section first introduces the principle of deep recurrent networks including simple recurrent neural network (RNN), Long Short-Term Memory (LSTM) \cite{Ref_LSTM}, and Gated Recurrent Unit (GRU) \cite{Ref_GRU}, followed by explaining how to apply a recurrent network to build a channel predictor \cite{Ref_Jiang_VTC2020}. The statistics of predicted CSI and the computational complexity for these predictors are also analyzed.
\subsection{Deep Recurrent Networks}

Unlike  unidirectional information flow in feed-forward neural networks,  RNN has recurrent self-connections to memorize historical information, exhibiting great potential in time-series prediction \cite{Ref_TimeSeriesPrediction}. The activation of the previous time step is fed back as part of the input for the current step. In a simple RNN, its $l^{th}$ recurrent layer is generally modeled as
\begin{equation}
\label{Eqn_RNN_hiddenlayer}
\mathbf{d}_t^{(l+1)}=\mathcal{R}^{(l)}(\mathbf{d}_t^{(l)})=\delta_h \left( \mathbf{W}^{(l)} \mathbf{d}_t^{(l
)} + \mathbf{U}^{(l)} \mathbf{d}_{t-1}^{(l+1)} +\mathbf{b}^{(l)} \right),
\end{equation}
where $\mathbf{W}^{(l)}$ and $\mathbf{U}^{(l)}$ are weight matrices of the $l^{th}$ layer, $\mathbf{b}^{(l)}$ is a bias vector,  $\mathbf{d}_t^{(l)}$ and $\mathbf{d}_{t}^{(l+1)}$ represent the input and output for layer $l$ at time $t$, respectively, $\mathbf{d}_{t-1}^{(l+1)}$ is the feedback from the previous step, $\mathcal{R}^{(l)}(\cdot)$ stands for the relation function for the input and output of the $l^{th}$ RNN hidden layer, and the activation function  often selects  the \emph{hyperbolic tangent} denoted by $\mathrm{tanh}$, \textcolor{red}{i.e.,} $\delta_h(x) =(e^{2x}-1)/(e^{2x}+1)$.

\begin{figure}[!hbpt]
\centering
\includegraphics[width=0.48\textwidth]{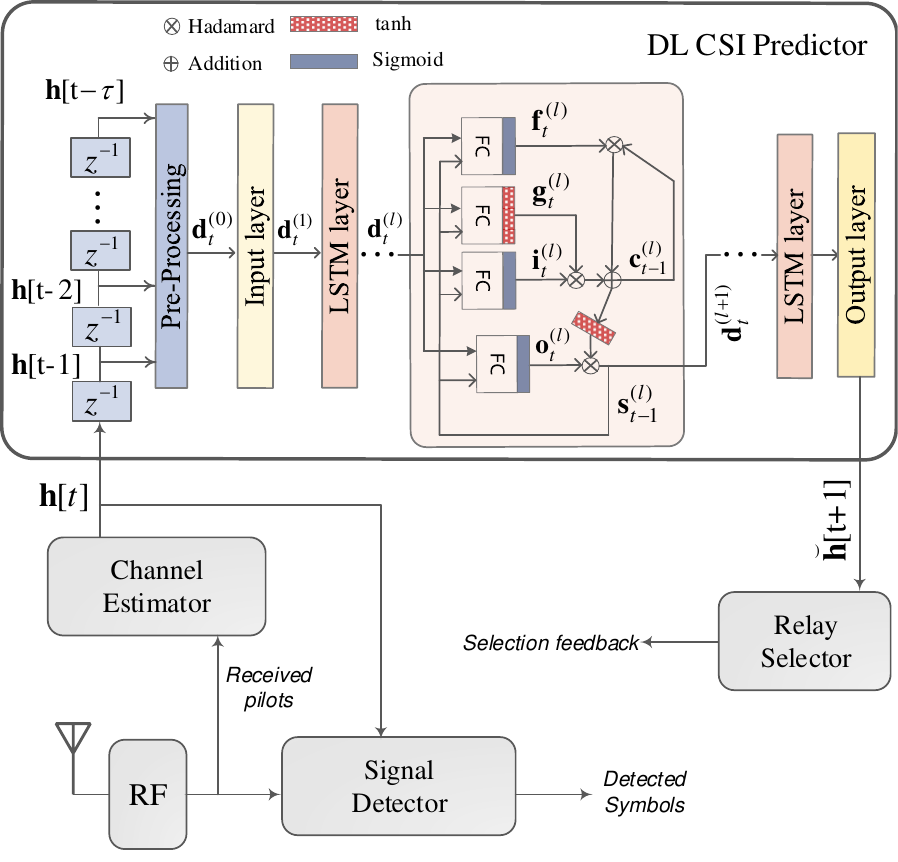}
\caption{Block diagram of the receiver integrated a DL-based channel predictor that mainly consists of an input layer, an output layer, and $L$ hidden layers. The $l^{th}$ hidden layer is opened to detail the internal structure of an LSTM memory block and its information flow. To remain historical channel information, a tapped-delay line is applied to form a series of consecutive CSI samples for the input layer. The predictor is inserted between the channel estimator and relay selector, transforming measured CSI to predicted CSI \textit{transparently} without any other modifications for an ORS system.}
\label{Figure_DLpredictor}
\end{figure}

Using typical stochastic gradient descent (SGD) method to train a recurrent network, the back-propagated error signals tend to zero that implies a prohibitively-long convergence time. To tackle this gradient-vanishing problem, Hochreiter and Schmidhuber proposed Long Short-Term Memory in their pioneer work of \cite{Ref_LSTM}, which introduced  \textit{cell} and \textit{gate} into the RNN structure. The former is a special memory unit and the latter regulates read and write access to the cell. In 1999, Gers \emph{et al.} \cite{Ref_LSTM_forget} further introduced a new gate that learns to reset the hidden state at appropriate times.  Then, a common LSTM cell has three gates: an \emph{input} gate controlling the extent of new information flows into the cell, a \emph{forget} gate  to filter out useless memory, and an \emph{output} gate that controls the extent to which the memory is applied to generate the activation.
The upper part of \figurename \ref{Figure_DLpredictor} shows the graphical depiction of a deep LSTM network consisting of an input layer, $L$ hidden layers, and an output layer.  Let's use the $l^{th}$ hidden layer as an example to shed light on how an activation signal goes through the network.  There are two hidden states - the short-term state $\mathbf{s}_{t-1}^{(l)}$ and the long-term state $\mathbf{c}_{t-1}^{(l)}$.  The input $\mathbf{d}_t^{(l)}$ and $\mathbf{s}_{t-1}^{(l)}$ jointly activate four fully connected (FC) layers, generating the activation vectors for the gates, i.e.,
\begin{equation} \left\{ \begin{aligned}
\label{Eqn_No1}
\mathbf{i}_t^{(l)} & = \delta_g \left( \mathbf{W}^{(l)}_{i}\mathbf{d}_t^{(l)} + \mathbf{U}_{i}^{(l)} \mathbf{s}_{t-1}^{(l)} +\mathbf{b}_i^{(l)} \right)\\
\mathbf{o}_t^{(l)}  &= \delta_g \left( \mathbf{W}_{o}^{(l)} \mathbf{d}_t^{(l)}  + \mathbf{U}_{o}^{(l)} \mathbf{s}_{t-1}^{(l)} +\mathbf{b}_o^{(l)} \right)\\
\mathbf{f}_t^{(l)} & = \delta_g \left( \mathbf{W}_{f}^{(l)} \mathbf{d}_t^{(l)} + \mathbf{U}_{f}^{(l)} \mathbf{s}_{t-1}^{(l)} +\mathbf{b}_f^{(l)} \right)
\end{aligned}, \right.
\end{equation}
where $\mathbf{W}$ and $\mathbf{U}$ are weight matrices for the FC layers, $\mathbf{b}$ represents bias, subscripts  $i$, $o$, and $f$ associate with the input, output, and forget gate, respectively, and $\delta_g$ stands for the logistic \emph{Sigmoid} function $\delta_g(x) = 1/(1+e^{-x})$.
The  current  long-term state  $\mathbf{c}_t^{(l)}$ is obtained by first throwing away outdated memory at the forget gate and then adding new information selected by the input gate, i.e., $
\mathbf{c}_t^{(l)}  =  \mathbf{f}_t^{(l)} \otimes \mathbf{c}_{t-1}^{(l)} +  \mathbf{i}_t^{(l)} \otimes  \mathbf{g}_t^{(l)}$,
where the operator $\otimes$ denotes the Hadamard product (element-wise multiplication) and $\mathbf{g}_t^{(l)}=\delta_h ( \mathbf{W}_{g}^{(l)} \mathbf{d}_t^{(l)} + \mathbf{U}_{g}^{(l)} \mathbf{s}_{t-1}^{(l)} +\mathbf{b}_g^{(l)} )$.
The output of this hidden layer is computed by
\begin{equation}
\label{Eqn_LSTMinputoutput}
 \mathbf{d}_t^{(l+1)}  = \mathcal{L}^{(l)} \left(\mathbf{d}_t^{(l)}\right)= \mathbf{o}_t^{(l)} \otimes \delta_h \left( \mathbf{c}_{t}^{(l)} \right),
\end{equation}
where $\mathcal{L}^{(l)}(\cdot)$ represents the input-output function for the $l^{th}$ LSTM layer. Note that the current short-term state is equal to the output, i.e., $  \mathbf{s}_t^{(l)}=\mathbf{d}_t^{(l+1)}$.

Despite of its short history, LSTM has achieved a great success and been commercially applied in many AI products such as Apple Siri and Google Translate. After its emergence,  the research community published a number of its variants, among which GRU proposed by Cho \textit{et al.} in \cite{Ref_GRU} drew lots of attention.  It's a simplified version with fewer parameters, but it exhibits even better performance over LSTM on certain smaller  and less frequent datasets. To simplify the structure, a GRU memory cell has only a single hidden state, and the number of gates is reduced to two: the \emph{update} and \emph{reset} gate. The activation vector for the update gate is computed by $\mathbf{z}_t^{(l)}  = \sigma_g ( \mathbf{W}_{z}^{(l)} \mathbf{d}_t^{(l)} + \mathbf{U}_{z}^{(l)} \mathbf{s}_{t-1}^{(l)} +\mathbf{b}_z^{(l)} )$, which decides the extend to which the memory content from the previous state will remain in the current state.
The reset gate controls whether the previous state is ignored, and when it tends to $0$, the hidden state is reset with the current input. It is given by $\mathbf{r}_t^{(l)}  = \sigma_g ( \mathbf{W}_{r}^{(l)}\mathbf{d}_t^{(l)} + \mathbf{U}_{r}^{(l)} \mathbf{s}_{t-1}^{(l)} +\mathbf{b}_r^{(l)} )$. Likewise, the previous hidden state $ \mathbf{s}_{t-1}^{(l)}$ goes through the cell, drops outdated memory, and inserts some new content, generating the current hidden state, that is
\begin{align}\label{Eqn_No2}
\mathbf{s}_t^{(l)} & =   (1- \mathbf{z}_t^{(l)}) \otimes \mathbf{s}_{t-1}^{(l)}\\ \nonumber &+  \mathbf{z}_t^{(l)} \otimes \sigma_h \left( \mathbf{W}_{s}^{(l)}\mathbf{d}_t^{(l)} + \mathbf{U}_{s}^{(l)}( \mathbf{r}_t^{(l)} \otimes \mathbf{s}_{t-1}^{(l)}) +\mathbf{b}_s^{(l)} \right).
\end{align}
The hidden state is also equal to its output of this hidden layer, i.e., $\mathbf{d}_t^{(l+1)}  = \mathcal{G}^{(l)}(\mathbf{d}_t^{(l)})=\mathbf{s}_t^{(l)}$, where $\mathcal{G}^{(l)}(\cdot)$ denotes the input-output function.

\subsection{DL-based Channel Predictor}

To shed light on the principle of a DL-based predictor, as shown in \figurename \ref{Figure_DLpredictor}, the chain of signal reception at the receiver is demonstrated. A predictor is inserted between the channel estimator and the relay selector, transforming  measured CSI to  predicted CSI as the input for relay selection. It is transparent and therefore an ORS system can be smoothly upgraded to a PRS system without any other modifications. Here, we use the \emph{centralized} relay selection  as an example, where the CSI of all $\mathbb{RD}$ links at time $t$ denoted by $\boldsymbol{h}_{d}[t]=\left[ h_{1,d}[t],...,h_{K,d}[t]\right]^T$ is processed at the destination.  For the \emph{distributed} selection,  each relay requires only  local CSI $h_{k,d}[t]$, which is simpler to handle and is therefore straightforwardly applicable.
 As illustrated in \figurename \ref{Figure_DLpredictor}, the instantaneous CSI $\boldsymbol{h}_{d}[t]$ measured by the channel estimator is fed into the predictor. To remain a few historical information, a tapped-delay line is applied. A series of consecutive CSI samples from $\boldsymbol{h}_{d}[t-\tau]$ to $\boldsymbol{h}_{d}[t]$  is available for the DL predictor to generate a $D$-step prediction $\check{\boldsymbol{h}}_{d}[t+D]$. 

As we know, a complex-valued fading coefficient can be expressed in polar form as $h_{k,d}[t]=a_{k,d}[t]e^{j\theta_{k,d}[t]}$, where $a_{k,d}[t]$ and $\theta_{k,d}[t]$ denote the magnitude and phase, respectively. Because the selection relies on the value of SNR, only the knowledge of magnitude $a_{k,d}[t]$ is enough, rather than complex-valued $h_{k,d}[t]$, which in turn can simplify the implementation of the channel predictor by employing  a neural network with real-valued weights and biases.
A pre-processing layer is in charge of adapting the format of CSI data to the input layer. In this case, the magnitudes need to be extracted, e.g., $\boldsymbol{a}_{d}[t]=\left[ a_{1,d}[t],...,a_{K,d}[t]\right]^T$ from $\boldsymbol{h}_{d}[t]$. After that, the extracted data $\left \{\boldsymbol{a}_{d}[t-\tau], \boldsymbol{a}_{d}[t-\tau+1],\cdots,\boldsymbol{a}_{d}[t-1], \boldsymbol{a}_{d}[t]\right \}$ are multiplexed as an input vector, we have
\begin{equation} \label{Eqn_inputVector}
   \boldsymbol{d}_{t}^{(0)}=\left[ a_{1,d}[t-\tau],  a_{2,d}[t-\tau],\cdots, a_{K,d}[t] \right]^T,
\end{equation}
which contains $K\times (\tau+1)$ entries. Feeding this input  vector  into the input feed-forward layer obtains $\mathbf{d}^{(1)}_t=\delta_h ( \mathbf{W}^{(I)} \mathbf{d}^{(0)}_t +\mathbf{b}^{(I)} )$, where $\mathbf{W}^{(I)}$ and  $\mathbf{b}^{(I)}$ denote the weight matrix and bias vector of the input layer. The activation of the $1^{st}$ hidden layer is exactly $\mathbf{d}^{(1)}_t$, thus $\mathbf{d}_t^{(2)}=\mathcal{L}^{(1)}(\mathbf{d}_t^{(1)})$ is generated and forwarded to the $2^{nd}$ hidden layer, where $\mathcal{L}^{(1)}\left(\cdot\right)$ is defined in (\ref{Eqn_LSTMinputoutput}). The activation goes through the network until the output layer gets the predicted CSI $\check{\boldsymbol{a}}_{d}[t{+}1]=\left[\check{a}_{1,d}[t{+}1],...,\check{a}_{K,d}[t{+}1] \right]^T$ (assuming $D=1$). It is computed by $\check{\boldsymbol{a}}_{d}[t{+}1]=\delta_h ( \mathbf{W}^{(O)} \mathbf{d}_t^{(L)} +\mathbf{b}^{(O)} )$, where $\mathbf{W}^{(O)}$ and  $\mathbf{b}^{(O)}$ denote the weight matrix and bias vector of the output layer, and the activation of the last hidden layer equals to $\mathbf{d}_t^{(L)}=\mathcal{L}^{(L)}( \ldots \mathcal{L}^{(2)}(\mathcal{L}^{(1)}(\mathbf{d}_t^{(1)})))$. The building of a deep recurrent network is flexible, for example, we can apply a hybrid network consisting of  RNN, GRU, and LSTM layers, like $\mathbf{d}_t^{(L)}=\mathcal{G}^{(L)} ( \ldots \mathcal{L}^{(2)} (\mathcal{R}^{(1)}(\mathbf{d}_t^{(1)})))$.

In addition to predict the magnitude of CSI, deep learning also provides the capability of processing complex-valued CSI \cite{Ref_Jiang_VTC2020B}. Instead of applying a deep neural network with complex-valued weights, which is currently not well supported by AI algorithms and software tools, we can decompose a fading coefficient into two
real numbers namely $h=\Re(h)+j\Im(h)$, where $\Re(\cdot)$ and $\Im(\cdot)$ take the real and imaginary units of a complex number, and the imaginary unit $j^2=-1$. Transforming $\textbf{h}_d[t]$ into $\textbf{c}_d[t]=\left[\Re(h_{1,d}[t]), ..., \Re(h_{K,d}[t]),\Im(h_{1,d}[t]), ..., \Im(h_{K,d}[t])  \right ]^T$ and training the predictor with such transformed CSI data,  the prediction output is $\check{\textbf{c}}_d[t+1]$ when feeding $\textbf{c}_d[t]$ at time $t$.  The complex-valued prediction $\check{\textbf{h}}_d[t{+}1]$ is obtained simply by taking a reverse manipulation over $\check{\textbf{c}}_d[t+1]$.

\subsection{Statistics of Predicted CSI}
To analyze the performance of the proposed scheme, the statistics of predicted CSI is mandatory. When training a DL-based predictor, the objective is set to generate predicted CSI $\check{h}$ that approximates to the actual CSI as close as possible.  It is therefore assumed that $\check{h}$ has the same distribution as $h$ and  follows zero-mean complex Gaussian distribution, i.e., $
\check{h} {\sim} \mathcal{CN}(0, \sigma_{\check{h}}^2)$.
Then, the instantaneous SNR $\gamma_{A,B}$ conditioned on its predicted version $\check{\gamma}_{A,B}{=}|\check{h}_{A,B}|^2 P_A /\sigma_n^2$ follows non-central Chi-square distribution with two degrees of freedom, whose Probability Density Function (PDF) is
\begin{IEEEeqnarray}{ll}
\label{Eqn_conditioned} \nonumber
f_{\gamma_{_{A,B}}|\check{\gamma}_{_{A,B}}}(\gamma|\check{\gamma}) = & \\  \frac{1}{\bar{\gamma}_{_{A,B}}(1-\rho^2)} e^{-\frac{\gamma+\rho^2\check{\gamma}}{\bar{\gamma}_{_{A,B}}(1-\rho^2)}} I_0\left( \frac{2\rho\sqrt{ \gamma \check{\gamma}}}{\bar{\gamma}_{_{A,B}}(1-\rho^2)} \right),  &
\end{IEEEeqnarray}
where $I_0(\cdot)$ denotes the $zero\mathrm{th}$ order modified Bessel function of the first kind, and $\rho$ stands for the correlation coefficient between $\check{h}$ and $h$, like (\ref{Eqn_CorCoeff}), defined as $\rho= \mathbb{E}[h\check{h}^*]/{\sqrt{\mathbb{E}[|h|^2] \mathbb{E}[|\check{h}|^2]}}$.

\subsection{Computational Complexity}
In the context of cooperative diversity, the computational complexity mainly arises from multi-relay coordination and synchronization \cite{Ref_SYN02}. The simplicity of ORS is achieved thanks to single-relay transmission that substantially lowers the amount of signalling overhead among multiple relays. A direct comparison of different schemes is not easy and does not provide real insight. That is why most of the works in this field  \cite{Ref_OPR, Ref_OPR2, Ref_Vicario, Ref_Seyfi, Ref_Torabi_Capacity01, Ref_Torabi, Ref_PartialRelaySelection,  Ref_Xiao, Ref_Jiang_GSC, Ref_Chen, Ref_Li_RelaySelection, Ref_MineTRANS} did not provide a quantitative analysis on complexity. On the other hand, the complexity of the proposed scheme comes mainly from the DL-based predictor, which is always a concern for the application of deep learning.  From a practical perspective, it is more meaningful to make clear its demand on computing resources  in comparison with the availability of off-the-shelf hardware. Hence, let's focus on assessing the complexity of the DL predictors in terms of floating-point operations per second (FLOPS).

A deep recurrent network can be quantitatively modelled as follows: an input layer with $N_i$  neurons, an output layer with $N_o$ neurons, and $L$ hidden layers, which has $N_h^{l}$ neurons at layer $l=1,\ldots,L$. To begin with the input layer, it computes $\delta_h ( \mathbf{W}^{(I)} \mathbf{d} +\mathbf{b}^{(I)} )$, where the matrix multiplication generates $N_iN_h^1$ floating-point multiplicative operations and  $(N_i-1)N_h^1$ additive operations, and the addition of the bias vector consumes $N_h^1$ operations, amounting to  a total of $O^i=2N_iN_h^1$. Note that the amount of computation raised by the activation function is negligible compared to the matrix multiplication, which is usually ignored in the calculation of complexity for deep learning. Likewise, it is easy to know that the output layer corresponds to $O^o=2N_h^LN_o$. For an RNN hidden layer as given in (\ref{Eqn_RNN_hiddenlayer}), the number of operations equals to $O^{l}=(2N_h^{l-1}-1)N_h^l+(2N_h^{l}-1)N_h^l+N_h^{l}$,
where the first term corresponds to the calculation of $ \mathbf{W}^{(l)} \mathbf{d}_t^{(l
)}  $, the second is for $ \mathbf{U}^{(l)} \mathbf{d}_{t-1}^{(l+1)}$, and the third is due to the addition of the bias.  For simplicity, $O^{l}$ can be approximated to $2N_h^{l-1}N_h^l+2(N_h^l)^2$. Then, the overall complexity for a simple RNN is given by
\begin{IEEEeqnarray}{ll}
\label{Eqn_complexity_RNN}O_{rnn}&=O^i+O^o+ \sum_{l=1}^L O^l\\ \nonumber
&\approx2\left[ N_iN_h^1+N_h^LN_o +\sum_{l=1}^L\left( N_h^{l-1}N_h^l+\left(N_h^l\right)^2 \right)  \right],
\end{IEEEeqnarray} where we apply $N_h^{0}=N_i$ for a simpler expression.
As derived from (\ref{Eqn_No1})-(\ref{Eqn_LSTMinputoutput}), the number of operations for the matrix multiplication on an LSTM layer is $4$ times that of an RNN layer, i.e., $4O^l$. The computation for the gate control, which has totally $7N_h^l-3$ operations, can be neglected. Therefore, the complexity of an LSTM network is approximated by
\begin{equation}
\label{Eqn_complexity_deepLSTM}
O_{lstm} \approx 2\left[N_iN_h^1+N_h^LN_o + \sum_{l=1}^{L} 4\left( N_h^{l-1}N_h^l+\left(N_h^l\right)^2 \right)\right].
\end{equation}
 Similarly, we can derive the expression for GRU, i.e.,
 \begin{equation}
 \label{Eqn_complexity_GRU}
O_{gru} \approx 2\left[N_iN_h^1+N_h^LN_o + \sum_{l=1}^{L} 3\left ( N_h^{l-1}N_h^l+\left(N_h^l\right)^2 \right)\right]
\end{equation}
Suppose all layers has identical number $n$ of neurons, we can further simplify (\ref{Eqn_complexity_RNN})-(\ref{Eqn_complexity_GRU}). As listed in Table \ref{table_complexity}, the complexity of recurrent networks is $\mathcal{O}(n^2)$, it is moderate if the number of neurons per layer is not too large.
\begin{table}[!h]
\renewcommand{\arraystretch}{1.3}
\caption{The complexity of deep recurrent networks.}
\label{table_complexity}
\centering
\begin{tabular}{|c|c|c|}
\hline
\textbf{Networks} & \textbf{Complexity per Step} &\textbf{FLOPS}\\
\hline \hline
 RNN & $4(1+L)n^2$ &$4(1+L)f_pn^2$  \\ \hline
GRU & $4(1+3L)n^2$& $4(1+3L)f_pn^2$ \\ \hline
LSTM & $4(1+4L)n^2$ & $4(1+4L)f_pn^2$  \\ \hline
\end{tabular}
\end{table}
During either the training phase using the typical SGD algorithm, or the predicting phase,  the required floating-point operations at each time step is identical. Consequently, (\ref{Eqn_complexity_RNN})-(\ref{Eqn_complexity_GRU}) are applicable for measuring the complexity of both training and prediction. Note that the above expressions are the complexity per step, we need to know the  frequency of prediction denoted by $f_p$, i.e., the number of steps performed per second,  to figure out FLOPS. Section VII will further discuss the complexity quantitatively after the values of these parameters are determined.

\section{Predictive Relay Selection}
\begin{figure}[!hbtp]
\centering
\includegraphics[width=0.45\textwidth]{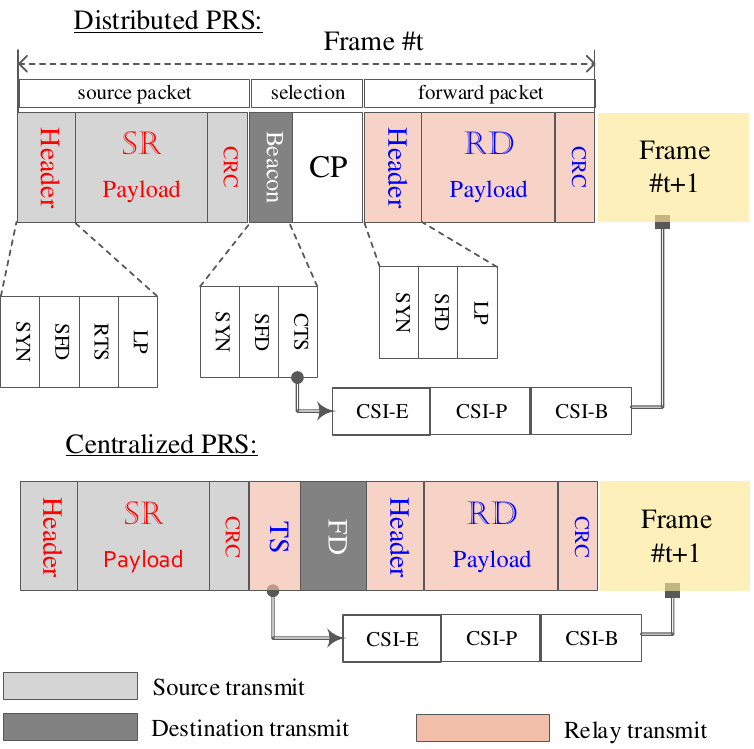}
\caption{Frame structure of the proposed scheme for the distributed (upper) and centralized (lower) relay-selection schemes. A frame is organized in three steps: a source packet for $\mathbb{SR}$ transmission, relay selection, and a forward packet for $\mathbb{RD}$ transmission. A packet consists of a header, payload, and a cyclic redundancy check (CRC) code. The header is a combination of some of the following fields: a synchronization (SYN) preamble - a sequence of known bits used for frequency offset correction and time alignment, start frame delimiter (SFD) - a pattern of bits applied to define the beginning of a packet,  pilot signals called Ready-To-Send (RTS), Clear-To-Send (CTS), or training sequence (TS), and length of payload (LP) representing the number of symbols in the payload. The instantaneous CSI of frame $t$ is measured through carrying out CSI-Estimation (CSI-E) and then CSI-Prediction (CSI-P) forecasts the possible CSI for frame $t+1$, which is buffered (CSI-B) and is fetched at the next frame for a timely relay selection.} 
\label{Figure_Implementation}
\end{figure}
Taking advantage of new degree of freedom opened by channel prediction, we propose the PRS scheme that intends to achieve high performance in fast time-varying channels while keeping up the full diversity in slow fading. The implementation of cooperative relay-selection schemes are mainly divided into two categories:  distributed \cite{Ref_OPR}  and centralized \cite{Ref_Seyfi}. The former relies on a timer at each relay, and applies a contention period (CP) to choose the best relay in a distributed manner. The latter has a centralized controller, e.g., the destination, which measures the CSI of all $\mathbb{RD}$ links and makes decision.  Instead of being immediately used to select the best relay for the current frame, the measured CSI is applied to generate predicted CSI for the next frame. Such a prediction horizon relaxes the tight requirement of time procedure and therefore provides the flexibility to design an advanced relaying strategy. Without loss of generality, as depicted in $\mathrm{Algorithm}~1$, we first depict an implementation example for the distributed PRS with the DF strategy, as follows:
\begin{enumerate}[\IEEEsetlabelwidth{5)}]
\item
At frame $t$, as illustrated in \figurename \ref{Figure_Implementation}, the source broadcasts a packet consisting of a header, payload, and a CRC code. Other nodes (the relays and destination) synchronize with the source by means of the SYN preamble.  Relay $k$, $k \in \{1,\cdots,K \}$ measures its local CSI $h_{s,k}[t]$  by estimating RTS, which is used   to detect received data symbols. Those relays that correctly decode the source's signal (i.e., passing CRC checking) comprise a $\mathcal{DS}$.
\item
A beacon containing CTS is sent from the destination, so that relay $k$ can estimate $h_{d,k}[t]$ and then $h_{k,d}[t]$ is known  due to channel reciprocity. It feeds $h_{k,d}[t]$ into its  \textit{local} channel predictor to generate $\check{h}_{k,d}[t+1]$, and buffers it for its usage at the upcoming frame $t+1$.
\item
Meanwhile, relay $k$ belonging to $\mathcal{DS}$ fetches $\check{h}_{k,d}[t]$  that was buffered at the previous frame $t-1$. This operation starts once the beacon arrives, in parallel with Step $2$.
\item
Each relay starts a timer with a duration inversely proportional to the magnitude of CSI, e.g., $T_t \propto 1/|\check{h}_{k,d}[t]|$. It is possible that this duration is too long due to a very small channel gain. To deal with this anomaly, a maximal duration $T_m$ is added.
\item
The timer on the relay with the largest channel gain expires first, and then it sends a flag packet to announce\footnote{Due to the ``hidden" node problem,  signal propagation delay, and the switch time from receive to transmit mode in a transceiver, the probability of having two or more relay timers expire within an uncertainty interval is nonozero, causing  transmission collision among ``best" relays. The detail analysis of  collision probability  refers to Section III of \cite{Ref_OPR}.  }.
\item
Once received the best relay's notification, other relays flush their timers and keep silent. The selected relay forwards the signal until the end of this frame.
\end{enumerate}

The frame structure for the distributed PRS shown in \figurename \ref{Figure_Implementation}  is also suitable for AF relaying networks. Only three main modifications are required: the best relay is determined in terms of $\min(|\check{h}_{s,k}[t]|,|\check{h}_{k,d}[t]|)$ rather than $|\check{h}_{k,d}[t]|$ in the DF relaying, the best relay only amplifies the received signal without detection, and therefore CRC is not needed, as detailed in $\mathrm{Algorithm}~2$.

\begin{algorithm}
\label{Alg_1}
\caption{Distributed DF PRS }
\begin{algorithmic}
\FOR{$t=1,2,...$}
\STATE $s$ sends RTS
\STATE $s$ sends data payload $\boldsymbol{x}[t]$
\WHILE{$k=1,...,K$}
\STATE estimate $h_{s,k}[t]$
\STATE detect: $\hat{\boldsymbol{x}}[t]=f(\boldsymbol{y}_{s,k}[t],h_{s,k}[t])$
\IF {$\hat{\boldsymbol{x}}[t]$ is error-free}
\STATE fetch $\check{h}_{k,d}[t]$ from Buffer
\STATE start a timer $\left(T_t\propto\frac{1}{|\check{h}_{k,d}[t]|}\right)$ $\cap$ $\left( T_t\leqslant T_m\right)$
\ENDIF
\ENDWHILE
\STATE $d$ sends CTS
\STATE $\dot{k}=\arg \max_{k\in \mathcal{DS}}\left(|{\check{h}_{k,d}[t]}|\right)$ sends a flag
\STATE $\dot{k}$ transmits $\hat{\boldsymbol{x}}[t]$
\WHILE{$k=1,...,K$}
\STATE estimate $h_{k,d}[t]$
\STATE predict and buffer $\check{h}_{k,d}[t+1]$
\ENDWHILE
\ENDFOR
\end{algorithmic}
\end{algorithm}
\begin{algorithm}
\caption{Distributed AF PRS }
\begin{algorithmic}
\FOR{$t=1,2,...$}
\STATE $s$ sends RTS
\STATE $s$ sends $\boldsymbol{x}[t]$
\WHILE{$k=1,...,K$}
\STATE estimate $h_{s,k}[t]$
\STATE predict and buffer  $\check{h}_{s,k}[t+1]$
\ENDWHILE
\STATE $d$ sends CTS
\WHILE{$k=1,...,K$}
\STATE fetch $\check{h}_{s,k}[t]$, $\check{h}_{k,d}[t]$ from Buffer
\STATE start a timer $T_t\propto \frac{1}{\min(|\check{h}_{s,k}[t]|,|\check{h}_{k,d}[t]|)}$ $\cap$ $\left( T_t\leqslant T_m\right)$
\STATE estimate $h_{k,d}[t]$
\STATE predict and buffer $\check{h}_{k,d}[t+1]$
\ENDWHILE
\STATE $\dot{k}=\arg \max_{k}\left( \min(|\check{h}_{s,k}[t]|,|\check{h}_{k,d}[t]|) \right)$ sends a flag
\STATE $\dot{k}$ transmits $\boldsymbol{y}_{\dot{k}}[t]$
\ENDFOR
\end{algorithmic}
\end{algorithm}

Moreover, the proposed scheme is also applicable to cooperative networks with centralized relay selection.  Its centralized version for DF relays is depicted as follows:
\begin{enumerate}[\IEEEsetlabelwidth{5)}]
\item
At frame $t$, as illustrated in \figurename \ref{Figure_Implementation}, the source broadcasts a packet containing a header, payload, and a CRC code. The relays achieve synchronization via the SYN preamble, estimate RTS to get the local CSI, and detect received data symbols.
\item
Once the termination of the $\mathbb{SR}$ transmission, the relays send out their respective TSs simultaneously.
\item
The destination can estimate the CSI of all $\mathbb{RD}$ links, i.e., $\boldsymbol{h}_{d}[t]=\left[ h_{1,d}[t],...,h_{K,d}[t]\right]^T$, if the TSs are orthogonal. Feeding $\boldsymbol{h}_{d}[t]$ into the \textit{global} predictor at the destination,  $\check{\boldsymbol{h}}_{d}[t{+}1]$ is obtained and then buffered for the usage at the next frame. Note that only a \textit{global} predictor  is needed within a cooperative network in contrast to the distributed PRS where each relay has a local predictor.
\item
Meanwhile, the destination fetches the predicted CSI $\check{\boldsymbol{h}}_{d}[t]=\left[ \check{h}_{1,d}[t],...,\check{h}_{K,d}[t]\right]^T$ that is  buffered at the previous frame $t{-}1$. This operation starts once the TSs arrive, in parallel with Step $3$.
\item
The destination selects the best relay in terms of $\dot{k}=\arg \max_{k}\left(|{\check{h}_{k,d}[t]}|\right)$ and the selection decision is fed back (FD) to the relays.
\item
The selected relay checks whether it correctly detects data symbols in the source packet by checking CRC. If yes, it relays the signal in the forward packet. Otherwise, it sends non-acknowledgement to trigger a relay re-selection process or the termination of this frame (whose data will be re-transmitted at the next framework \cite{Ref_zhao2005practical}).
\end{enumerate}
The centralized PRS using the DF strategy is also described in $\mathrm{Algorithm}~3$, while its AF counterpart can be derived, as from $\mathrm{Algorithm}~1$ to $\mathrm{Algorithm}~2$, and therefore is not repeated here because of the page limitation.

\begin{algorithm}
\caption{Centralized DF PRS }
\begin{algorithmic}
\FOR{$t=1,2,...$}
\STATE $s$ sends RTS
\STATE $s$ sends $\boldsymbol{x}[t]$
\WHILE{$k=1,...,K$}
\STATE send $k^{th}$ TS
\STATE estimate $h_{s,k}[t]$
\STATE detect: $\hat{\boldsymbol{x}}[t]=f(\boldsymbol{y}_{s,k}[t],h_{s,k}[t])$
\ENDWHILE
\STATE $d$ fetch $\check{\boldsymbol{h}}_{d}[t]$ from Buffer
\STATE select $\dot{k}=\arg \max_{k}\left(|{\check{h}_{k,d}[t]}|\right)$ and feed back
\IF {$\hat{\boldsymbol{x}}[t]$ on $\dot{k}$ is error-free}
\STATE $\dot{k}$ transmits $\hat{\boldsymbol{x}}[t]$
\ELSE
\STATE $d$ re-selects $\dot{k}$ or terminates
\ENDIF
\STATE $d$ estimates $\boldsymbol{h}_{d}[t]$ from TS
\STATE predict and buffer $\check{\boldsymbol{h}}_{d}[t+1]$
\ENDFOR
\end{algorithmic}
\end{algorithm}

\section{Outage Probability Analysis}
The performance of PRS will be analyzed with respect to (w.r.t.) outage probability and channel capacity, which are  key performance indicators to assess cooperative diversity techniques. In this section, we first derive the closed-form formulas of outage probabilities for DF and AF PRS, respectively, and then get their capacity expressions in the following section.
\subsection{Outage Probability for DF PRS}

In the Information Theory \cite{Ref_Tse}, the \emph{outage} points to the event that instantaneous channel capacity falls below a target rate $R$, where reliable communication is not achievable whatever channel coding used. The metric to measure the probability of outage is referred to as outage probability that is defined as $P(R){=} \mathbb{P} \left\{ \log_2(1+ \gamma ) < R \right\}$, where $\mathbb{P}$ is the notation of mathematical probability. In the DF relaying, the number of relays in a decoding subset varies from time to time due to the channel fading. Let's categorize all decoding subsets containing $M$ relays  into one group denoted by $\mathcal{DS}_M$, $M=0,1,\ldots,K$. These $M$ relays are probably different, namely $M$ out of $K$ relays, resulting in $ \binom{K}{M}$  combinations. In other words, $\mathcal{DS}_M$ is a set of decoding subsets, i.e., $\mathcal{DS}_M{=}\left\{\mathcal{DS}_M^p\left|p{=}1,...,\binom{K}{M}\right.\right\}$,  where $\mathcal{DS}_M^p$ denotes the $p^{th}$ element of $\mathcal{DS}_M$.
Then, the outage probability of PRS with DF relays can be calculated by
\begin{equation}
\label{Eqn_outageProb_first}
P_{prs}^{DF}(R)=\sum_{M=0}^{K} \sum_{p=1}^{\binom{K}{M}} \mathbb{P} (R|\mathcal{DS}_M^p) \mathbb{P} (\mathcal{DS}_M^p),
\end{equation}
where $\mathbb{P}(\mathcal{DS}_M^p)$ is the occurrence probability of $\mathcal{DS}_M^p$, and $\mathbb{P} (R|\mathcal{DS}_M^p)$ is the outage probability conditioned on $\mathcal{DS}_M^p$. Suppose that all $\mathbb{SR}$ links are independent and identically-distributed (\emph{i.i.d.}), the values of $\mathbb{P} (\mathcal{DS}_M^p)$ for any $p{\in}\left\{1,...,\binom{K}{M}\right\}$ are equal, and as well $\mathbb{P}(R|\mathcal{DS}_M^p)$ if all $\mathbb{RD}$ channels are \emph{i.i.d}. Then, (\ref{Eqn_outageProb_first}) can be simplified to
\begin{equation}
\label{Eqn_outageProb_first_simp}
P_{prs}^{DF}(R)=\sum_{M=0}^{K} \mathbb{P} \left(R||\mathcal{DS}|=M\right) \mathbb{P} \left(|\mathcal{DS}|=M\right),
\end{equation}
where $|\cdot|$ represents the cardinality of a set and $\mathbb{P}(|\mathcal{DS}|{=}M)$ denotes the probability that the number of relays in a decoding subset is $M$. With Rayleigh fading, the instantaneous SNR of each $\mathbb{SR}$ channel is exponentially distributed, whose Cumulative Distribution Function (CDF) is given by
\begin{equation}
\label{Eqn_cdf}
F_{\gamma_{s,k}}(\gamma)=1-e^{-\frac{\gamma}{\bar{\gamma}_{s,k}}}, \:\:\: \: \gamma>0.
\end{equation}
According to (\ref{Eqn_DS}), the probability that a relay correctly decodes the received signal,  or $\gamma_{s,k} \geqslant \gamma_o$, equals to $1{-}F_{\gamma_{s,k}}(\gamma_o)$. $M$ out of $K$ relays falling into the $\mathcal{DS}$ follows the binomial distribution, we would obtain
\begin{equation}
\label{Formular_nchoosek}
\mathbb{P}(|\mathcal{DS}|=M) = \binom{K}{M} \left( e^{-\frac{\gamma_o}   {\bar{\gamma}_{s,k}}}\right)^M \left(1-e^{-\frac{\gamma_o }{\bar{\gamma}_{s,k}}} \right)^{K-M}.
\end{equation}
By far, the second term in (\ref{Eqn_outageProb_first_simp}) is determined. Let's turn to the first term $\mathbb{P} \left(R||\mathcal{DS}|=M\right)$, which  is derived,  conditioned on the number of $M$,  as follows:
\subsubsection{$M=0$} If no relay can successfully decode the original signal, the signal transmission
fails, i.e.,
\begin{equation}
\label{Eqn_L0}
\mathbb{P} (R||\mathcal{DS}|=0)=1.
\end{equation}
\subsubsection{$M=1$} Only one relay correctly decodes the signal, it acts as the best relay directly without selection. Similar to (\ref{Eqn_cdf}), we obtain the CDF of the received SNR for this $\mathbb{RD}$ link as $F_{\gamma_{\dot{k},d}}(\gamma){=}1-e^{-\gamma/\bar{\gamma}_{k,d}}$, resulting in
\begin{equation}
\label{Eqn_L1}
\mathbb{P} (R||\mathcal{DS}|=1) =F_{\gamma_{\dot{k},d}}(\gamma_o)=1-e^{-\frac{\gamma_o }{\bar{\gamma}_{k,d}}} .
\end{equation}

\subsubsection{$M{>}1$}
In this case, the best relay is opportunistically selected from the $\mathcal{DS}$ in terms of predicted CSI, that is $\dot{k}= \arg\max_{k\in\mathcal{DS}}\left( \check{\gamma}_{k,d}\right)$.
To simplify the derivation, we use $\mathcal{A}_{\dot{k}}$ to represent the event of $\check{\gamma}_{\dot{k}}=\max_{k\in\mathcal{DS}}\left( \check{\gamma}_{k,d}\right)$.
The predicted CSI is applied only for relay selection, whereas the post-processing SNR during signal transmission should be the actual SNR $\gamma_{\dot{k}}$, whose CDF can be calculated by
\begin{equation}
\label{Eqn_CDF}
F_{\gamma_{\dot{k}}}(\gamma)  =\sum_{\dot{k}=1}^M \mathbb{P}(\gamma_{\dot{k}} \leqslant \gamma| \mathcal{A}_{\dot{k}}) \mathbb{P} \left(\mathcal{A}_{\dot{k}} \right),
\end{equation}
where $\mathbb{P} (\mathcal{A}_{\dot{k}})$ denotes the occurrence probability of $\mathcal{A}_{\dot{k}}$. Under the assumption of \emph{i.i.d} channels,   each relay in the decoding subset has the same chance to get the largest SNR, thus $\mathbb{P} (\mathcal{A}_{\dot{k}})=1/M$. Besides, $\mathbb{P}(\gamma_{\dot{k}} \leqslant \gamma| \mathcal{A}_{\dot{k}}) $ is the probability that the actual SNR is below an arbitrary threshold $\gamma$ conditioned on $\mathcal{A}_{\dot{k}}$, which is computed by
\begin{equation}
\label{Eqn_conditionalP}
\mathbb{P}(\gamma_{\dot{k}} \leqslant \gamma| \mathcal{A}_{\dot{k}})  =\int_{0}^{\gamma}  \int_{0}^{\infty}  f_{\gamma_{\dot{k}}|\check{\gamma}_{\dot{k}}}(\gamma|\check{\gamma}) f_{\check{\gamma}_{\dot{k}}| A_{\dot{k} }}(\check{\gamma}) d \gamma d\check{\gamma},
\end{equation}
where $f_{\gamma_{\dot{k}}|\check{\gamma}_{\dot{k}}}(\gamma|\check{\gamma})$ stands for the PDF of $\gamma_{\dot{k}}$ conditioned on its predicted version $\check{\gamma}_{\dot{k}}$, as given in (\ref{Eqn_conditioned}), and $f_{\check{\gamma}_{\dot{k}}| A_{\dot{k} }}(\check{\gamma})$ denotes the PDF of $\check{\gamma}_{\dot{k}}$ in the case of $A_{\dot{k} }$. Analogue to the multi-user selection with a max-SNR scheduler in \cite{Ref_multiuserSelectionP}, we have
\begin{equation}
\label{Eqn_pdf_cond_1}
f_{\check{\gamma}_{\dot{k}}| A_{\dot{k} }}(\check{\gamma})= \frac{M e^{-\frac{\check{\gamma}}{\bar{\gamma}_{k,d}}}}{\bar{\gamma}_{k,d}} \left ( 1-e^{-\frac{\check{\gamma}}{\bar{\gamma}_{k,d}}}\right)^{M-1}.
\end{equation}
Equation (\ref{Eqn_conditionalP}) is solved given (\ref{Eqn_conditioned}) and (\ref{Eqn_pdf_cond_1}), and then substituting $\mathbb{P}(\gamma_{\dot{k}} \leqslant \gamma| \mathcal{A}_{\dot{k}})$ into (\ref{Eqn_CDF}), we have
\begin{equation}
\label{Eqn_cdf_final}
F_{\gamma_{\dot{k}}}(\gamma)=  \sum_{m=0}^{M-1} \binom{M-1}{m} \frac{(-1)^m}{m+1} \left ( 1-e^{-\frac{\gamma(m+1)}{\bar{\gamma}_{k,d} \left[1+m (1-\rho^2 )\right]}  }\right).
\end{equation}
Thus, the conditional outage probability at $M>1$ is
\begin{equation}
\label{Eqn_L2}
\mathbb{P} (R||\mathcal{DS}|=M) =F_{\gamma_{\dot{k}}}(\gamma_o).
\end{equation}
Substituting (\ref{Formular_nchoosek}), (\ref{Eqn_L0}), (\ref{Eqn_L1}), and (\ref{Eqn_L2}) into (\ref{Eqn_outageProb_first_simp}), the closed-form expression of outage probability for DF PRS is obtained:
\begin{IEEEeqnarray}{lll}
\label{Eqn_PORS_outCSI}
\nonumber
  P_{prs}^{DF}(\gamma_o)&=&\left(1-e^{-\frac{\gamma_o }{\bar{\gamma}_{s,k}}} \right)^K  \\ \nonumber
  &+& \sum_{M=1}^{K}   \sum_{m=0}^{M-1} \binom{M-1}{m} \frac{(-1)^m}{m+1} \left ( 1-e^{\frac{-\gamma_o(m+1)}{\bar{\gamma}_{k,d} \left[1+m (1-\rho^2 )\right]}  }\right) \\
  &\cdot & \binom{K}{M} \left( e^{-\frac{\gamma_o}   {\bar{\gamma}_{s,k}}}\right)^M \left(1-e^{-\frac{\gamma_o }{\bar{\gamma}_{s,k}}} \right)^{K-M}.
\end{IEEEeqnarray}

\subsection{Outage Probability for AF PRS}
In the AF relaying, the best relay is selected in terms of the equivalent end-to-end CSI. The predicted SNR of the best relay is the largest, i.e., $\check{\gamma}_{\dot{k}} = \max_{k\in[1,...,K]} \left\{\min ( \check{\gamma}_{s,k}, \check{\gamma}_{k,d} )\right\}$.
However, the calculation of outage probability requires the PDF of the actual SNR, rather than the predicted SNR, i.e.,
 \begin{equation}
\label{Eqn_AF_OutageProb_PDF}
P_{prs}^{AF}(\gamma_o){=} \int_0^{\gamma_o} f_{\gamma_{\dot{k}}}(\gamma) d\gamma,
\end{equation}
where $\gamma_o$ is the threshold SNR defined in (\ref{Eqn_DS}).
Conditioned on its predicted version $\check{\gamma}_{\dot{k}}$, the PDF of $\gamma_{\dot{k}}$ is computed by
\begin{equation}
\label{Eqn_AF_overall_PDF}
f_{\gamma_{\dot{k}}}(\gamma)= \int_0^{\infty}  f_{\gamma_{\dot{k}}|\check{\gamma}_{\dot{k}}}(\gamma|\check{\gamma}) f_{\check{\gamma}_{\dot{k}}}(\check{\gamma}) d\check{\gamma},
\end{equation}
where $f_{\check{\gamma}_{\dot{k}}}(\check{\gamma}) $ stands for the PDF of $\check{\gamma}_{\dot{k}}$. Under the assumption of \emph{i.i.d.} Rayleigh fading, we can first figure out its CDF as
\begin{equation}
\label{Eqn_CDF_AF}
F_{\check{\gamma}_{\dot{k}}}(\check{\gamma}) = \mathbb{P}(\check{\gamma}_{\dot{k}}<\check{\gamma}) = \prod_{k=1}^K \mathbb{P}(\check{\gamma}_{k}<\check{\gamma}) = \prod_{k=1}^K F_{\check{\gamma}_{k}}(\check{\gamma}).
\end{equation}
Since $\check{\gamma}_{s,k}$ and $\check{\gamma}_{k,d}$ are exponentially distributed, $\check{\gamma}_k=\min ( \check{\gamma}_{s,k}, \check{\gamma}_{k,d} )$ also follows the exponential distribution with a mean of $
\bar{\gamma}_e=\frac{\bar{\gamma}_{s,k} \bar{\gamma}_{k,d}}{\bar{\gamma}_{s,k} + \bar{\gamma}_{k,d}}
$. Like (\ref{Eqn_cdf}), we have $F_{\check{\gamma}_{k}}(\check{\gamma}){=}1-e^{-\check{\gamma}/\bar{\gamma}_e}$ and then
 (\ref{Eqn_CDF_AF}) gets solved as
\begin{equation}
F_{\check{\gamma}_{\dot{k}}}(\check{\gamma}) = \left(
1-e^{-\check{\gamma}/\bar{\gamma}_e}\right)^K.
\end{equation}
Taking its derivative, yields
\begin{equation}
\label{Fomular_Pdf}
f_{\check{\gamma}_{\dot{k}}}(\check{\gamma}) = \frac{\partial F_{\check{\gamma}_{\dot{k}}}(\check{\gamma})}{\partial \check{\gamma}}=\frac{K}{\bar{\gamma}_e} e^{-\check{\gamma}/\bar{\gamma}_e} \left[
1-e^{-\check{\gamma}/\bar{\gamma}_e}\right]^{K-1}.
\end{equation}
For the sake of mathematical tractability,  according to \cite{Ref_analyticalapproach}, (\ref{Fomular_Pdf}) is transformed into another form as
\begin{equation}
\label{Eqn_AF_PDF_bestSNR_set}
f_{\check{\gamma}_{\dot{k}}}(\check{\gamma}) = \sum_{k=1}^{K} \binom{K}{k} \frac{(-1)^{(k-1)}k}{\bar{\gamma}_e} e^{-\frac{k\check{\gamma}}{\bar{\gamma}_e}}.
\end{equation}
Substituting  (\ref{Eqn_conditioned}) and (\ref{Eqn_AF_PDF_bestSNR_set}) into (\ref{Eqn_AF_overall_PDF}), yields
\begin{IEEEeqnarray}{ll}
\label{Eqn_Outage_AF_Pdf}
f_{\gamma_{\dot{k}}}(\gamma) &= \sum_{k=1}^{K} \binom{K}{k} \frac{(-1)^{(k-1)}k}{\bar{\gamma}_e} e^{-\frac{k\gamma}{\bar{\gamma}_e}} \\ \nonumber
& \times \int_0^{\infty}  e^{-\check{\gamma}\left( \frac{\rho^2}{(1-\rho^2)\bar{\gamma}_e} + \frac{k}{\bar{\gamma}_e} \right) } I_0\left( \frac{2\sqrt{\rho^2 \gamma \check{\gamma}}}{\bar{\gamma}_{e} (1-\rho^2) } \right)d\check{\gamma} .
\end{IEEEeqnarray}
Applying Eq. (6.614.3) of \cite{Ref_Book_Integrals}, i.e., $\int_0^{\infty} e^{-\alpha x} I_0(2\sqrt{\beta x})dx =\frac{1}{\alpha} e \left(\frac{\beta}{\alpha}\right)$, with $\alpha=\left( \frac{\rho^2}{(1-\rho^2)\bar{\gamma}_e} + \frac{k}{\bar{\gamma}_e} \right)$ and $\beta=\frac{\rho^2 \gamma }{[\bar{\gamma}_{e} (1-\rho^2)]^2}$, we would solve (\ref{Eqn_Outage_AF_Pdf}) as
\begin{equation}
\label{Eqn_AF_PDF_bestSNR}
f_{\gamma_{\dot{k}}}(\gamma) = \sum_{k=1}^{K} \binom{K}{k} \frac{(-1)^{(k-1)}k}{\bar{\gamma}_e[k(1-\rho^2)+\rho^2]} e^{-\frac{k \gamma}{[k(1-\rho^2)+\rho^2]\bar{\gamma}_e}}.
\end{equation}
Substituting (\ref{Eqn_AF_PDF_bestSNR}) into (\ref{Eqn_AF_OutageProb_PDF}), the analytical expression of outage probability for AF PRS can be figured out, i.e.,
\begin{equation}
\label{Eqn_AF_outageProb}
P_{prs}^{AF}(\gamma_o) = \sum_{k=1}^{K} \binom{K}{k} (-1)^{k} \left[e^{-\frac{k \gamma_o}{[k(1-\rho^2)+\rho^2]\bar{\gamma}_e}} -1\right].
\end{equation}

\section{Capacity Analysis}
Channel capacity is another key performance metric, indicating the maximal transmission rate, at which data can be delivered over a wireless channel with negligible error probability. In general, it can be calculated by taking the integral of the received SNR's PDF, namely $C=\int_0^{\infty} \log_2(1+\gamma)f(\gamma)d\gamma$. In the context of cooperative relaying, a closed-form expression of channel capacity is usually hard to derive. For instance, in \cite{Ref_Torabi_Capacity01}, the final expression still contains an exponential integral $\int_1^\infty t^{-1}e^{-\lambda t}dt$.  To avoid such intractability in the PDF-based analysis, we apply another approach taking advantage of Moment Generating Function (MGF) \cite{Ref_Jiang_VTC_MGF}, defined as  $M_{\gamma}(s){=}\mathbb{E}[e^{-s\gamma}]$. The MGF-based approach is depicted as follows:
\newtheorem{lemma}{Lemma}
\begin{lemma}
\label{Lemma_02}
The ergodic capacity of a wireless system can be derived through the MGF of the received SNR  \cite{Ref_MGFcapacity}, that is
\begin{equation}
\label{Eqn_ergocapactiy_approx}
C = \frac{1}{\ln(2)} \sum_{q=1}^Q w_q \Phi (s_q) \left[ \left. \frac{\partial}{\partial s} M_{\gamma}(s) \right|_{s \rightarrow s_q}  \right],
\end{equation}
where $\ln$ is the natural logarithm,  $Q$ stands for the number of iterations (truncated at $Q{=}200$ is already accurate enough), $\Phi(s)$ denotes a special mathematical function called Meijer's G, i.e., $
\Phi(s) = -G_{2,1}^{0,2} \left[ \cdot  \right]$,
the variable $s_q$ is a function of $q$, which is given by $s_q=\tan \left[ 0.25\pi\cos \left( (q-0.5)\pi/Q \right) + 0.25\pi \right]$, and another variable
\begin{equation}
w_q = \frac{\pi^2 \sin \left[ (q-0.5)\pi/Q \right]}{4Q\cos^2 \left\langle 0.25\pi \cos \left[ (q-0.5)\pi/Q\right] + 0.25\pi \right\rangle }.
\end{equation}
\end{lemma}

\begin{IEEEproof}
The derivation refers to Appendix $A$.
\end{IEEEproof}

\begin{figure*}[!htpb]
\normalsize
\begin{IEEEeqnarray}{lll}
\label{Eqn_DF_PRS_Capacity_Final} \nonumber
C_{prs}^{DF} & \: = \: &   \frac{1}{2 \ln(2)} \left\{  K\left ( e^{-\frac{\gamma_{_{o}}}{\bar{\gamma}_{s,k}}} \right) \left(1-e^{-\frac{\gamma_{_{o}} }{\bar{\gamma}_{s,k}}} \right)^{K-1}
   \sum_{q=1}^Q - \bar{\gamma}_{k,d} w_q \Phi(s_q)   \left( \frac{1}{1+s_q \bar{\gamma}_{k,d}} \right)^{2}  \right.    \\
 & \: + \: & \left.  \sum_{M=2}^{K}    \left ( e^{-\frac{  \gamma_{_{o}} }{\bar{\gamma}_{s,k}}} \right)^M \left(1-e^{-\frac{\gamma_{_{o}} }{\bar{\gamma}_{s,k}}} \right)^{K-M}
           \sum_{q=1}^Q w_q \Phi(s_q)  \sum_{m=0}^{M-1} \binom{M{-}1}{m} \frac{(-1)^{m+1} \bar{\gamma}_{k,d} \left[1+m (1-\rho^2 )\right] }{\left \langle m+1+s_q\bar{\gamma}_{k,d} \left[1+m (1-\rho^2 )\right]\right \rangle^2 }\right\}
\end{IEEEeqnarray}
\hrulefill
\vspace*{4pt}
\end{figure*}

\begin{figure*}[!htbp]
\normalsize
\begin{equation}
\label{Eqn_AF_Ergocapacity}
C^{AF}_{prs}  =   \frac{1}{\ln(2)} \sum_{q=1}^Q \sum_{k=1}^K (-1)^k   w_q \Phi(s_q)
 \binom{K}{k} \left( \frac{\bar{\gamma}_{e}k\left[k(1-\rho^2)+\rho^2\right]}{\left \langle k+s_q \bar{\gamma}_e  \left[k(1-\rho^2)+\rho^2\right]\right\rangle^2 } \right)
\end{equation}
\hrulefill
\vspace*{4pt}
\end{figure*}

\subsection{Capacity of DF PRS}
Analogous to (\ref{Eqn_outageProb_first_simp}), the channel capacity of the proposed scheme using DF relays can be computed by
\begin{equation}
\label{GS_capacity_OSTC}
C_{prs}^{DF} =\sum_{M=0}^{K} C^{M} \Pr(|\mathcal{DS}|{=}M),
\end{equation}
where $\Pr(|\mathcal{DS}|{=}M)$ is given in (\ref{Formular_nchoosek}), and $C^{M}$ denotes the capacity for the end-to-end channel conditioned on $|\mathcal{DS}|{=}M$, which is analyzed as follows:

\subsubsection{$M=0$} It means no relay can correctly decode the source's signal, leading to $C^0=0$.
\subsubsection{$M = 1$}
If only one relay is available, it directly serves as the best relay without the need of  selection. For a Rayleigh channel, the MGF of $\gamma_{\dot{k}}$ is given by
\begin{equation}
\label{Eqn_MGF_L1}
\mathcal{M}_{\gamma_{\dot{k}}}(s) = \frac{1}{1+s\bar{\gamma}_{k,d}}.
\end{equation}
Due to the half-duplex mode in dual-hop cooperative systems, the  capacity for the $\mathbb{EE}$ channel has to be halved, multiplying a factor of $1/2$. Substituting (\ref{Eqn_MGF_L1}) into (\ref{Eqn_ergocapactiy_approx}),  we have
\begin{equation}
\label{GS_capacity_L}
C^1 = \frac{1}{2\ln(2)} \sum_{q=1}^Q w_q \Phi (s_q)   \frac{-\bar{\gamma}_{k,d}}{(1+s_q\bar{\gamma}_{k,d})^2}.
\end{equation}

\subsubsection{$M > 1$}
The best relay is opportunistically selected from the $\mathcal{DS}$, and its PDF of the received SNR  $f_{\gamma_{\dot{k}}}(\gamma)$ can be obtained by taking the derivative of (\ref{Eqn_cdf_final}).
Upon this, we can derive the MGF of $\gamma_{\dot{k}}$ as follows
\begin{IEEEeqnarray}{ll}
\label{EQN_MGF001} \nonumber
\mathcal{M}_{\gamma_{\dot{k}}}(s)&= \int_0^{\infty}   e^{-s\gamma}  f_{\gamma_{\dot{k}}}(\gamma) d\gamma\\
&= \int_0^{\infty}   e^{-s\gamma}  \frac{\partial F_{ \gamma_{\dot{k}}}(\gamma)}{\partial \gamma} d\gamma\\ \nonumber
                       &=  \sum_{m=0}^{M-1} \binom{M-1}{m} \frac{(-1)^m}{m+1+s  \bar{\gamma}_{k,d} \left[1+m (1-\rho^2 )\right] }.
\end{IEEEeqnarray}
Substituting  (\ref{EQN_MGF001}) into (\ref{Eqn_ergocapactiy_approx}), the capacity of the $\mathbb{RD}$ channel $C^{M}_{r,d}$ is obtained. Analogous to (\ref{GS_capacity_L}), a factor  $1/2$ is multiplied due to the half-duplex model, yields the $\mathbb{EE}$ capacity of
\begin{equation}
\label{GS_capacity_LLN}
    C^M=\frac{1}{2}C^{M}_{r,d}.
\end{equation}

Looking back to (\ref{GS_capacity_OSTC}), the required terms $\Pr(|\mathcal{DS}|{=}M)$ and $C^{M}$ are available. This enables the following theorem:
\newtheorem{theorem}{Theorem}
\begin{theorem}
The end-to-end ergodic capacity for the proposed scheme using DF relays over \emph{i.i.d.} Rayleigh channels  is given in a closed form by  (\ref{Eqn_DF_PRS_Capacity_Final}).
\end{theorem}
\begin{IEEEproof}
Substituting (\ref{Formular_nchoosek}), (\ref{GS_capacity_L}), and (\ref{GS_capacity_LLN})  into (\ref{GS_capacity_OSTC}), yields (\ref{Eqn_DF_PRS_Capacity_Final}).
\end{IEEEproof}

\subsection{Capacity of AF PRS}

Given $f_{\gamma_{\dot{k}}}(\gamma)$ in (\ref{Eqn_AF_PDF_bestSNR}), the  MGF of the actual SNR is calculated by $
M_{\gamma_{\dot{k}}}(s)=\int_0^{\infty}   e^{-s\gamma} f_{\gamma_{\dot{k}}}(\gamma) d\gamma $, yielding
\begin{equation}
\label{Formular_MGF_bestrelay}
\mathcal{M}_{\gamma_{\dot{k}}}(s) = \sum_{k=1}^K \frac{(-1)^{k-1} \binom{K}{k} }{1+s\bar{\gamma}_e[k(1-\rho^2)+\rho^2]/k},
\end{equation}
whose derivative is
\begin{IEEEeqnarray}{lll}
\label{Formular_MGF_der} \nonumber
\frac{\partial M_{\gamma_{\dot{k}}}(s)}{\partial s}  &= & \sum_{k=1}^K (-1)^k \binom{K}{k}  \\
& \times &\left( \frac{k\bar{\gamma}_{e}\left[k(1-\rho^2)+\rho^2\right]}{\left \langle k+s \bar{\gamma}_e  \left[k(1-\rho^2)+\rho^2\right]\right \rangle ^2}  \right).
\end{IEEEeqnarray}
\begin{theorem}
The end-to-end ergodic capacity for the proposed scheme using AF relays over \emph{i.i.d.} Rayleigh channels  is provided in a closed form by  (\ref{Eqn_AF_Ergocapacity}).
\end{theorem}
\begin{IEEEproof}
Substituting (\ref{Formular_MGF_der}) into (\ref{Eqn_ergocapactiy_approx}), yields (\ref{Eqn_AF_Ergocapacity}).
\end{IEEEproof}

\section{Numerical results}
In this section, we first introduce the acquisition of CSI datasets for training deep recurrent networks, and clarify how to decide hyper-parameters to obtain high prediction accuracy. Monte-Carlo simulation is carried out to get numerical results w.r.t. outage probability and channel capacity, which are applied to corroborate the theoretical analysis and conduct performance comparison with the existing schemes. Moreover, the robustness, scalability, and  complexity of the proposed scheme are evaluated.

\subsection{CSI Datasets}
\begin{figure*}[!t]
\centerline{
\subfloat[]{
\includegraphics[width=0.42\textwidth]{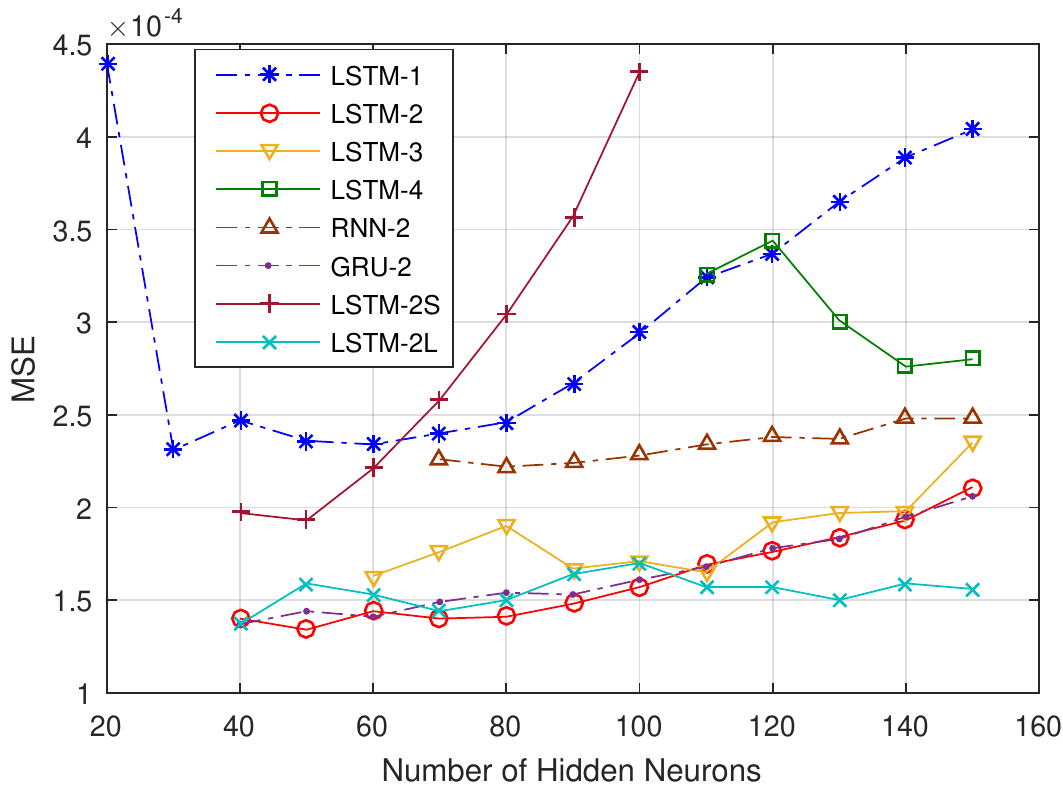}
\label{Fig_MSE}
}
\hspace{0mm}
\subfloat[]{
\includegraphics[width=0.42\textwidth]{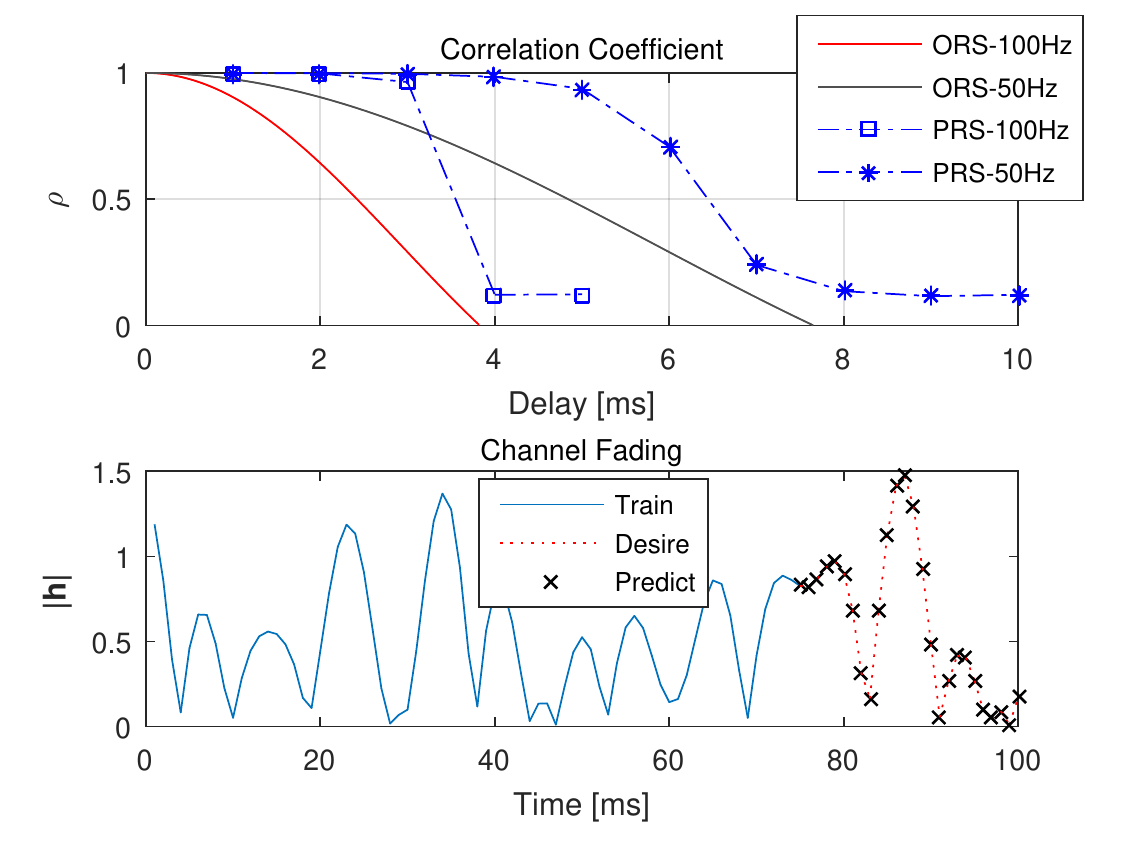}
\label{Fig_Corr}
}}
\hspace{15mm}
\caption{(a) Prediction accuracy with different hyper-parameters in terms of the number of hidden neurons; (b) The upper: Comparison of correlation coefficient for outdated and predicted CSI, and the lower: Illustration of a time-varying channel differentiating the training and predicting phase.}
\label{Fig_Result1}
\end{figure*}

A proper dataset is essential for training and testing a data-driven algorithm and plays a critical role to get high accuracy. We have ever established a wireless test-bed \cite{Ref_Jiang_ICCC_Testbed} based on the open-source 4G implementation, i.e., OpenAirInterface, to acquire realistic channel data.  Compared with the synthesis data, it shows no evident difference for the task of channel prediction. For simplicity, the simulation results provided in this section are obtained based on the synthesis  data acquired on MATLAB$^{\circledR}$ using its embedded wireless channel models. Following the channel assumption adopted by most of the previous works in this field,  we would apply single-antenna  flat-fading  \textit{i.i.d.} channels. Each channel follows the Rayleigh distribution with an average power gain of $0\mathrm{dB}$, where its fading coefficient $h$ is zero-mean circularly-symmetric complex Gaussian random variable with the variance of $1$, i.e., $h {\sim} \mathcal{CN}(0, 1)$.  To emulate fast fading environment, the  maximal Doppler shift is set to $f_d{=}100 \mathrm{Hz}$, which corresponds to a moving speed of around 100 \si[per-mode=symbol]{\kilo\meter\per\hour} at the carrier frequency of \SI{1}{\giga\hertz}. Continuous-time channel responses are sampled with a rate of $f_s{=}1 \mathrm{KHz}$, adhering to the assumption of flat fading, and therefore the interval of samples is  $T_s{=}1\mathrm{ms}$. Each channel generates a series of $10^6$ consecutive samples $ \{h[t]\left |  t{=}1,2,\ldots,10^6 \right.  \}$. The lower part of \figurename \ref{Fig_Corr} shows an example piece of such a channel.

\begin{figure*}[!t]
\centerline{
\subfloat[]{
\includegraphics[width=0.42\textwidth]{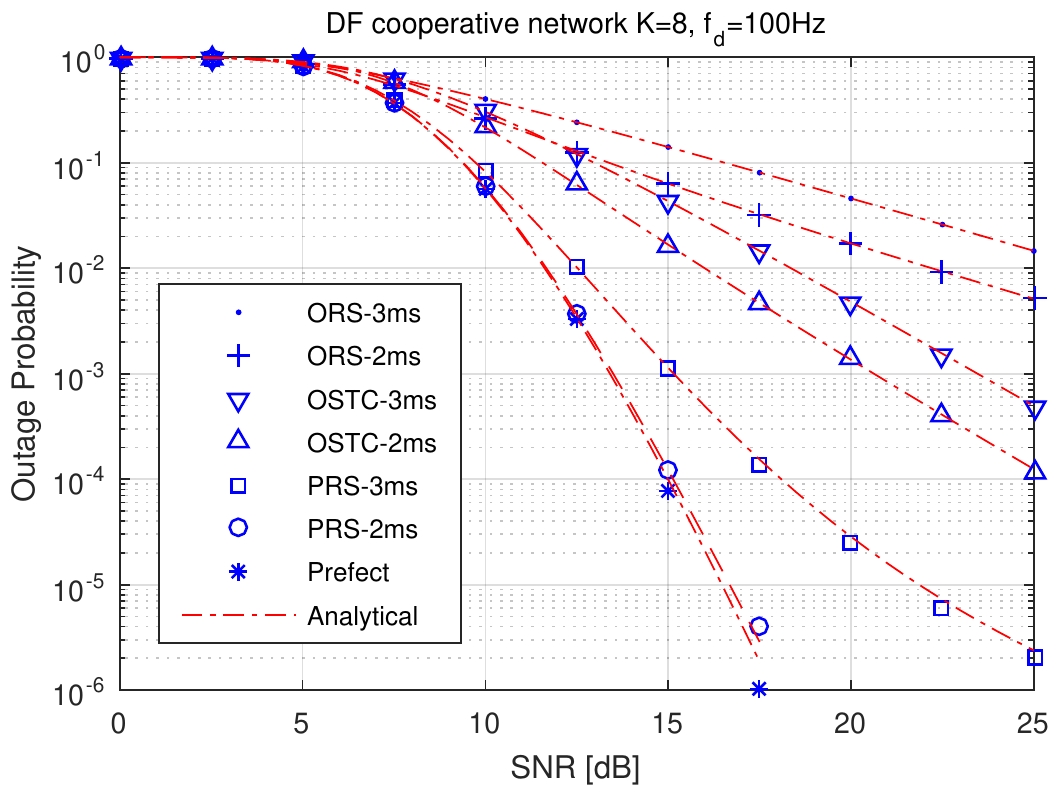}
\label{Fig_DF_outage}
}
\hspace{0mm}
\subfloat[]{
\includegraphics[width=0.42\textwidth]{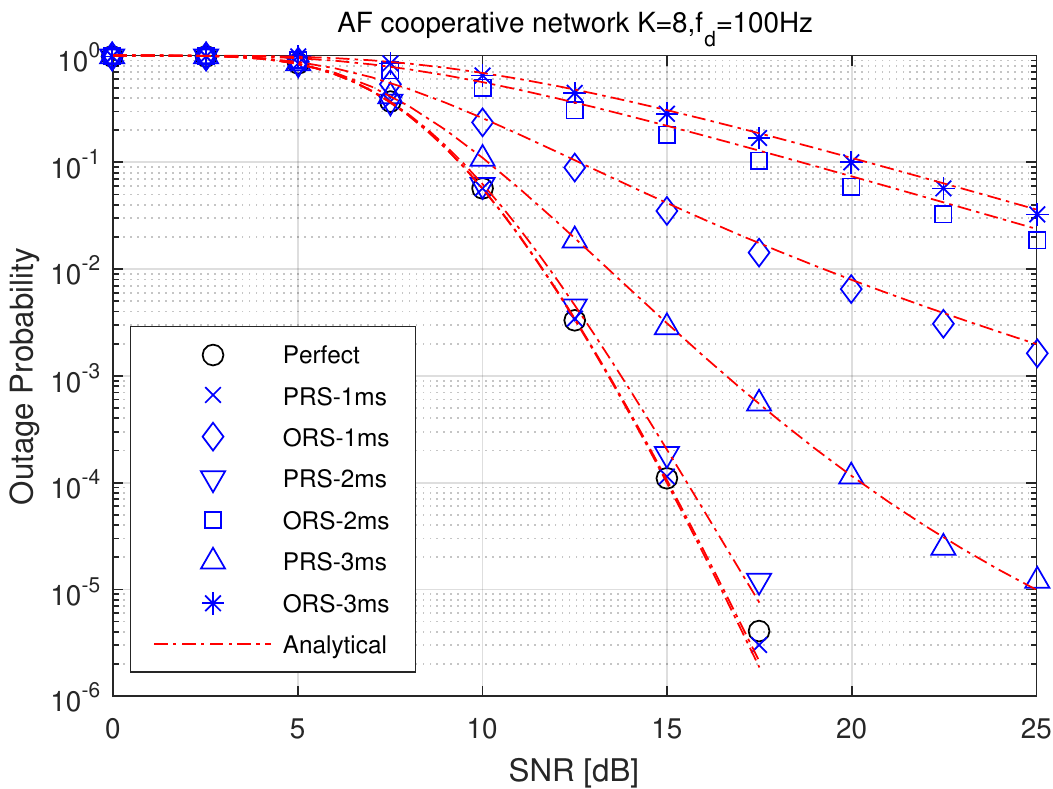}
\label{Fig_AF_outage}
}}
\hspace{15mm}
\caption{(a) Comparison of outage probability for ORS, OSTC, and PRS in a DF cooperative network with $K{=}8$ relays; (b) Comparison of outage probability for ORS and PRS in an AF cooperative network with $K{=}8$ relays.}
\label{Fig_Result2}
\end{figure*}

\subsection{Training the Predictor}
Hyper-parameters of a deep network, such as the number of layers or neurons, activation functions, training algorithms, and the length of training data, have a substantial impact on accuracy. It is worth clarifying how to tune a deep network on demand. A training process starts from an initial state where all weights and biases are randomly selected.  Using the centralized relay selection as an example, the input of the predictor at the destination is a magnitude vector $\mathbf{a}_d[t]$, while the output is its $D$-step-ahead prediction $\check{\mathbf{a}}_d[t{+}D]$. To measure prediction accuracy, mean squared error (MSE)  is applied as the cost function, namely $
\mathrm{MSE} = \frac{1}{T} \sum_{t=1}^ T \left \|  \check{\mathbf{a}}_d[t+D] - \mathbf{a}_d[t+D] \right \|^2$, where $T$ is the total number of channel samples for evaluation and $ \left \| \cdot \right \|$ notates the Frobenius norm of a vector.  Using the \emph{batch} training, a batch of $256$ samples is fed into the network per step. The output is compared with the desired values and the resultant error signals are propagated back through the network to update the weights by means of training algorithms such as the Adam optimizer \cite{Ref_kingma2017adam} used in our simulation. After a total of $10$ epochs, the trained network is employed to predict CSI.

\begin{table}[!bh]
\renewcommand{\arraystretch}{1.3}
\caption{Simulation configuration}
\label{table_SimParameters}
\centering
\begin{tabular}{c|c} 
\hline
\textbf{Parameters} & \textbf{Values}  \\
\hline \hline 
Sampling rate  &  $f_s=1000\mathrm{Hz}$ \\ \hline
Max. Doppler shift  &  $f_d=100\mathrm{Hz}$  \\ \hline
Channel model  & Rayleigh fading \\   \hline
Doppler spectrum & Jakes's model \\ \hline
Number of Relay & $K=8$  \\  \hline
Dataset size & $10^6$  \\   \hline \hline
Deep learning & LSTM netwok ($L=2$, $N_l=25$)\\ \hline
Training algorithm  & Adam optimizer \cite{Ref_kingma2017adam} \\   \hline
Batch size & 256  \\   \hline
Tapped-delay line & $\tau=4$ \\ \hline
Cost function & MSE \\ \hline
Actuation function  & $\mathrm{tanh}$  \\ \hline
\end{tabular}
\end{table}

\figurename \ref{Fig_MSE} compares the prediction accuracy of the predictors with respect to different hyper-parameters. Without loss of generality, we select a cooperative network with $K=8$ relays as the default scenario for simulation. The number of relays does not affect the superiority of the proposed scheme, which will be verified in the following part. The length of the tapped-delay line is selected according to the coherence time because too `\emph{old}' CSI samples are uncorrelated and do not provide useful information. As an example, we select $\tau=4$ for $f_d=100\mathrm{Hz}$, which is verified as the optimal setting in simulation.  The input vector defined in (\ref{Eqn_inputVector}) is thus $\textbf{d}_t^{(0)}=[a_{1,d}[t-4],a_{2,d}[t-4],\ldots,a_{8,d}[t]]^T$, which has a dimension of $K \times (\tau+1)=40$. One-step-ahead prediction $\check{\textbf{a}}_d[t+1]=[\check{a}_{1,d}[t+1],...,\check{a}_{8,d}[t+1]]^T$ is the output of the predictor.  Let's first look at the impact of the number of layers and the number of neurons. Starting from an LSTM network with  a single hidden layer, denoted by \emph{LSTM-1} in the legend of the figure, its accuracy curve  as a function of the number of hidden neurons likes an `U' shape. That is because the network suffers from the \emph{under-fitting} problem with only $20$ neurons in the hidden layer, while the \emph{over-fitting} problem appears at the turn point of $80$ neurons.  To make a fair comparison, the horizontal axis represents the total number of hidden neurons, which are evenly allocated to different layers. For instance, the point of `60' in the horizontal axis means a 2-hidden-layer network with $30$ neurons at either layer (denoted by \emph{LSTM-2}), a 3-hidden-layer network with $20$ neurons per layer (denoted by \emph{LSTM-3}),  or  a single layer with $60$ hidden neurons.  No matter how many neurons used in its single hidden layer, \emph{LSTM-1} cannot reach the high accuracy achieved by \emph{LSTM-2} and \emph{LSTM-3},  justifying the benefit of deep learning. But it does not mean that the more layers, the better,  as demonstrated by the worse result of \emph{LSTM-4}, which has 4 hidden layers. After known that 2-hidden-layer is the best choice for LSTM, we further observe the recurrent networks with 2 RNN or GRU hidden layers, indicated by \emph{RNN-2} and \emph{GRU-2}, respectively. As we can see, GRU performs as good as LSTM, whereas RNN is weak. Furthermore, we check the impact of the length of training data. The aforementioned results are measured with the default length of $5,000$ channel samples. We first shorten it to $2,500$, as shown by \emph{LSTM-2S}, the deep network seems to be under-fitted, leading to an obvious loss. On the contrary, if the length is doubled to $10,000$, as shown by \emph{LSTM-2L}, the performance keeps good and is even better with $110\sim150$ neurons. As a result, we select a 2-hidden-layer LSTM network with $25$ neurons at either layer and a training length of $5,000$, upon which the numerical results in the following figures are derived.

\subsection{Performance Comparison}
Numerical results of outage probability and channel capacity for PRS, ORS, and OSTC in the presence of perfect, predicted, and outdated CSI are obtained from Monte-Carlo simulation. As usual, an $\mathbb{EE}$ target rate of $R{=}1 \mathrm{bps/Hz}$ is applied for outage calculation. The total transmit power $P$ is equally allocated between two phases, where the source's power is $P_s{=}0.5P$, resulting in an  average SNR $\bar{\gamma}_{s,k}{=}0.5P/\sigma_n^2$, while $\bar{\gamma}_{k,d}{=}0.5P/\sigma_n^2$ for the $\mathbb{RD}$ link. In these figures, the numerical results are marked by \textbf{markers}, while the analytical results are plotted into \textbf{curves}. It can be seen that the markers fall into their corresponding curves, corroborating our theoretical analyses in the previous sections of this article.

\begin{figure*}[!t]
\centerline{
\subfloat[]{
\includegraphics[width=0.42\textwidth]{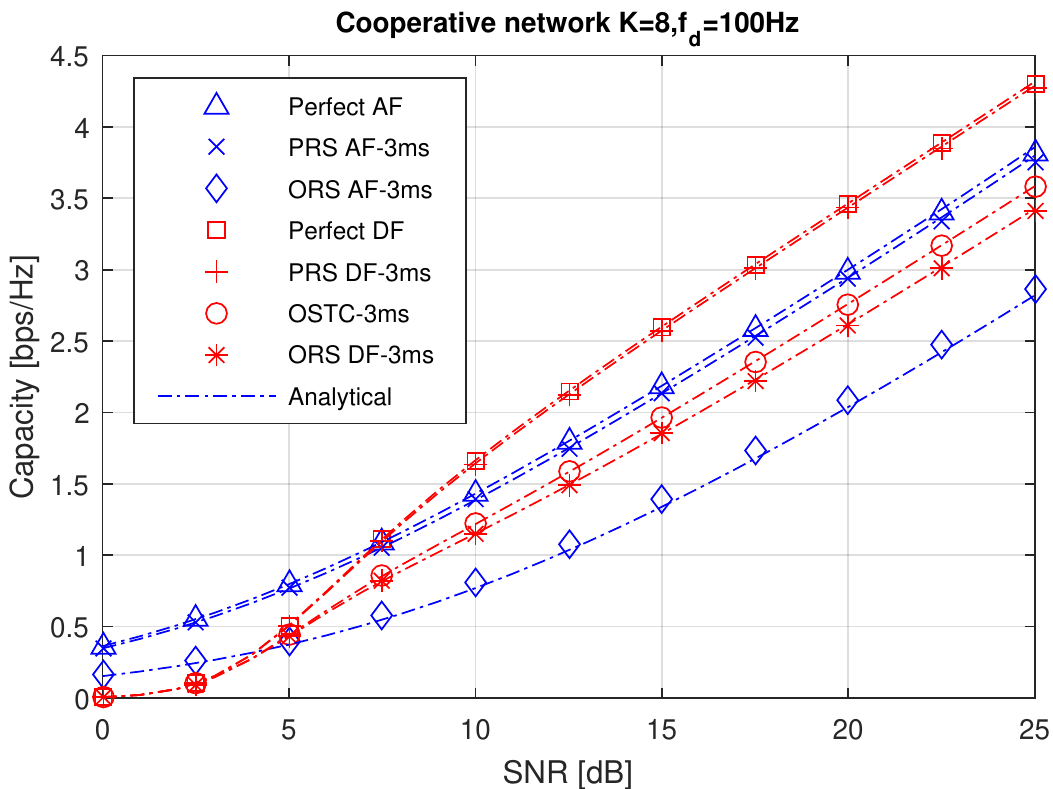}
\label{Fig_capacity}
}
\hspace{0mm}
\subfloat[]{
\includegraphics[width=0.42\textwidth]{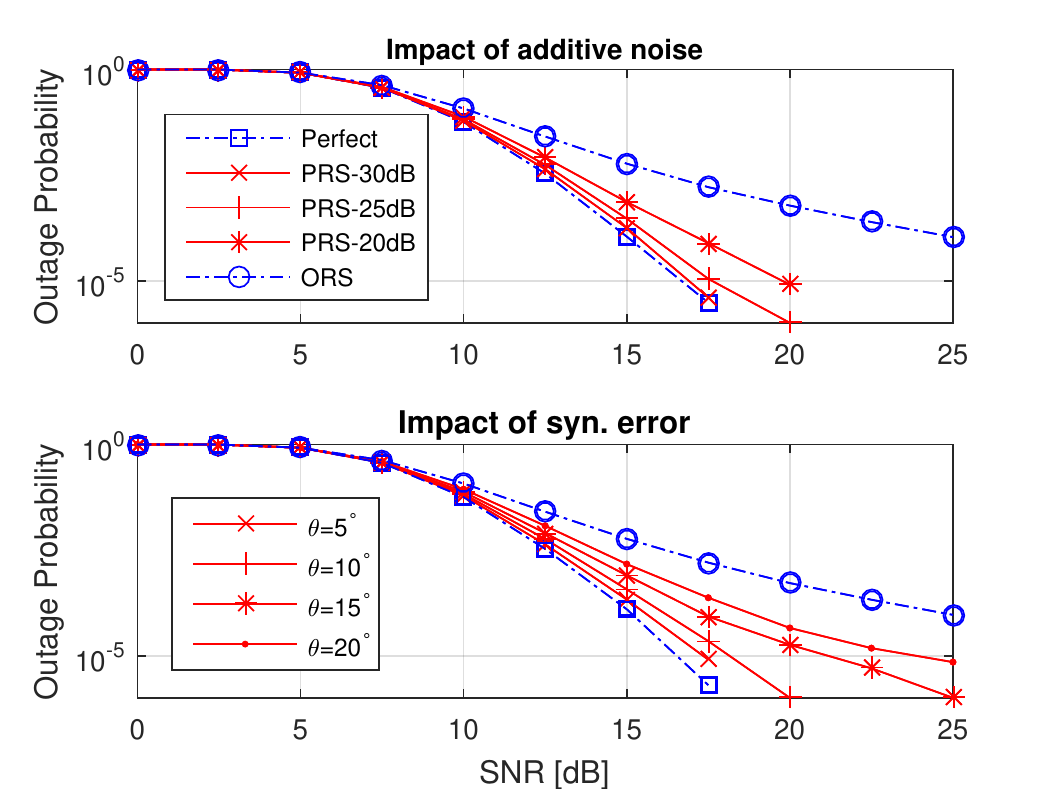}
\label{Fig_Impaire}
}}
\hspace{15mm}
\caption{(a) Comparison of channel capacity for ORS, OSTC, and PRS in a cooperative network with $K{=}8$ relays; (b) The impact of additive noise (the upper) and synchronization error (the lower) on the performance of outage probability.}
\label{Fig_Result3}
\end{figure*}

The performance of a cooperative network is directly affected by the quality of applied CSI. Let's first give a glance at the quality superior of predicted CSI. The correlation coefficient of outdated CSI  is calculated by (\ref{Eqn_Jakes}), e.g.,  $\rho_o=J_0(200\pi \tau)$ for $f_d=100\mathrm{Hz}$.  With the increase of delay $\tau$, the similarity between the outdated and actual CSI falls off, as indicated by $\mathrm{ORS{-}100Hz}$ in \figurename \ref{Fig_Corr}, until it becomes totally uncorrelated at nearly $\tau=4\mathrm{ms}$. The predicted CSI has higher quality, e.g., $\rho>0.95$ at $\tau =3\mathrm{ms}$, in comparison with $\rho_o \approx 0.29$ of the outdated CSI. At the point of $\tau=4\mathrm{ms}$, the quality of the predicted CSI suffers from a sudden drop, because the outdated CSI fed into the predictor is already uncorrelated and cannot provide any useful information about the actual CSI. It implies that the maximal prediction horizon is limited by the coherence time of a fading channel. Looking at the case of $50\mathrm{Hz}$, the predicted CSI also has  remarkably better quality over the outdated CSI.

Next, we compare the outage performance of three DF relaying schemes in a cooperative network with $K{=}8$ relays, as illustrated in \figurename \ref{Fig_DF_outage}. The relay selection with the perfect knowledge of CSI (i.e., $\rho{=}1$) is used as the benchmark, which has the diversity order of $8$ and decays at a rate of $1/\bar{\gamma}^8$, where $\bar{\gamma}=P/\sigma_n^2$ is the average $\mathbb{EE}$ SNR. With the delay of $\tau=2$ and $3\mathrm{ms}$, the quality of outdated CSI drops to $\rho_o=J_0(0.4\pi)\thickapprox 0.6425$ and $J_0(0.6\pi)\thickapprox0.2906$, respectively, which substantially deteriorates the performance. The diversity of ORS falls into $1$, i.e., no diversity, and the curve decays slowly  at a rate of $1/\bar{\gamma}$ in the high SNR regime.    OSTC can redeem some loss and achieve the diversity order of $2$ by using a pair of relays,  but its gap to the benchmark is still large, more than $7\mathrm{dB}$ at the level of $10^{-3}$. Making use of channel prediction, the quality of CSI can be improved to $\rho>0.95$. The proposed scheme achieves nearly the optimal performance with the horizon of $2\mathrm{ms}$ (by setting $D=2$ steps prediction), and remarkably outperforms OSTC with a gain of approximately $8\mathrm{dB}$ in the case of $3\mathrm{ms}$.
Additionally, \figurename \ref{Fig_AF_outage} compares the performance of ORS and PRS in a cooperative network with $K{=}8$ AF relays (OSTC is only applicable to DF relaying due to its utilization of space-time coding). With the horizon/delay of $1ms$, the proposed scheme receives the optimal performance, whereas ORS suffers from a loss of around $12\mathrm{dB}$ at the level of $10^{-3}$. Increased $\tau$ to $2\mathrm{ms}$ and $3\mathrm{ms}$, PRS substantially outperforms ORS with a gain of around $15\mathrm{dB}$.  The analytical results of PRS are given by (\ref{Eqn_PORS_outCSI}) and (\ref{Eqn_AF_outageProb}), which are corroborated by the numerical results, and those of ORS and OSTC are from \cite{Ref_MineTRANS}. 

Channel capacities for different schemes are comparatively provided in \figurename \ref{Fig_capacity}. Looking first at the AF relaying, ORS suffers from a capacity loss of around $1\mathrm{bps/Hz}$ if $\tau=3\mathrm{ms}$, but PRS can achieve a near-optimal capacity of $3\mathrm{bps/Hz}$ at the SNR of $\bar{\gamma}{=}20\mathrm{dB}$. In a DF cooperative network, the advantage of the proposed scheme is also obvious. For instance, ORS, OSTC, and PRS achieves $2.6$, $2.75$, and $3.5\mathrm{bps/Hz}$, respectively, where the capacity of PRS closely approaches to the perfect one. As we can see, the numerical results tightly agree with the curves of analytical results, validating the correctness of (\ref{Eqn_DF_PRS_Capacity_Final}) and (\ref{Eqn_AF_Ergocapacity}).

\subsection{Robustness}
In addition to its performance, the robustness of the proposed scheme against additive noise, synchronization error, mobility, and different fading statistics is evaluated. Noise is unavoidable during the process of acquiring CSI data, so it is necessary to make clear its impact on the performance.  Setting the received SNR of the pilot signals used to estimate  CSI to $30\mathrm{dB}$, as shown in \figurename \ref{Fig_Impaire}, the performance loss is negligible.  The impact gradually becomes clear when the SNR is decreased to $25\mathrm{dB}$ and $20\mathrm{dB}$, but it still obviously outperforms ORS. Unlike ordinary data delivery, the acquisition of CSI data dedicated to the training of deep networks can use more transmission resources, e.g., higher transmit power for pilots. Hence, an SNR of $30\mathrm{dB}$ or even higher is practically expected and the proposed scheme can be regarded as robust enough in the front of noise. Then, the effect of synchronization error  between two communicating nodes is also studied. If the maximal residual phase error is $\theta=5^{\circ}$, the loss is not evident compared to the perfect CSI. With a growing value of $\theta$, the performance deteriorates, but PRS with $\theta=20^{\circ}$ is still better than ORS. Unlike multi-relay transmission, PRS is a kind of single-relay transmission where MCFO and MTO is not required.  Its synchronization is as simple as that of point-to-point single-antenna communication link, which is a mature technique, and therefore keeping a residual error under $5^{\circ}$ is  achievable in practice.

\begin{figure}[!hbtp]
\centering
\includegraphics[width=0.45\textwidth]{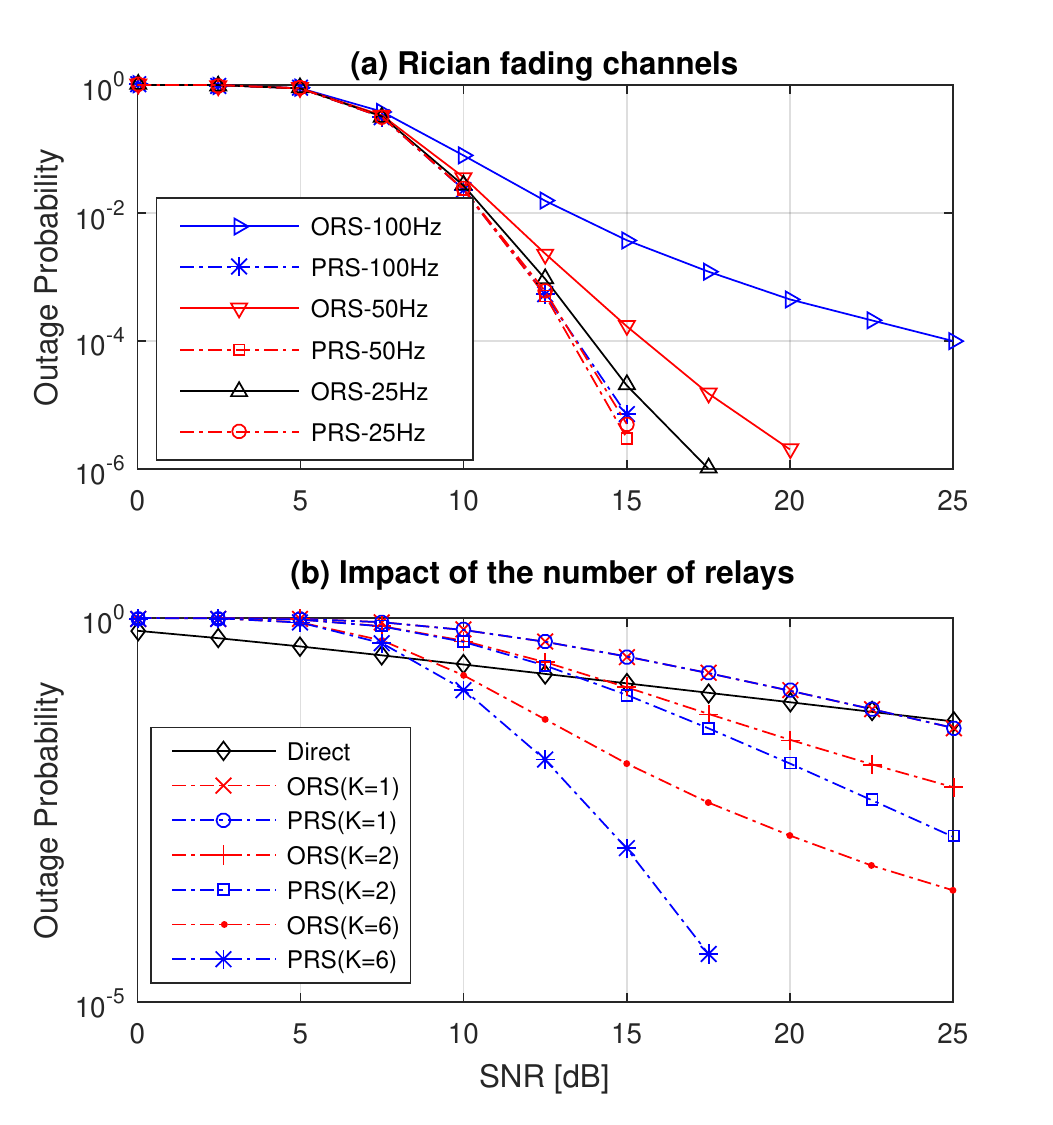}
\caption{ (a) The impact of moving speed (measured by the Doppler shift) on outage probability of ORS and PRS in a DF cooperative network with $K=8$ relays over \textit{Rician} fading channels (b) Outage probability of a DF cooperative network with $K=1,2$, and $6$ relays over Rayleigh fading channels, in comparison with direct transmission between the source and the destination. In the simulation, the source of DT transmits using power $P$ and full time duration $T$, in contrast to $P/2$ and $T/2$ used by the source in the relaying.} 
\label{Figure_others}
\end{figure}

 As a data-driven technique, a DL predictor treats a fading channel as a black box and only needs local channel measurements. The proposed scheme is therefore applicable for any kind of wireless channel statistics, isolated from radio propagation parameters such as fading distribution, the number of propagated paths, and the angle of arrival. It is interesting to examine the performance in Rician fading, where a dominating signal path exists between two communicating nodes, in addition to a large number of reflecting paths in Rayleigh fading.
\figurename \ref{Figure_others}a demonstrates the impact of mobility on outage probability over \textit{Rician} fading channels. Normalized by carrier frequency, the moving speed of a node is measured by the Doppler shift, as $25\mathrm{Hz}$, $50\mathrm{Hz}$, and $100\mathrm{Hz}$ used in the figure. At high speed ($f_d=\SI{100}{\hertz}$ corresponding to a velocity of $108\mathrm{km/h}$ at the carrier frequency of $1\mathrm{GHz}$), PRS shows much better performance over ORS. Such a superiority weakens with a slowdown of moving speed until the ORS scheme achieves the full diversity when measured CSI is not outdated. It proves our argument that the proposed scheme remains the full diversity in slow fading as same as the ORS scheme, and substantially outperforms in fast fading.

We also observe the impact of network scale on the performance of a DF cooperative network, as illustrated in \figurename \ref{Figure_others}b, in comparison with  direct transmission (DT). If there is only one relay available in the network ($K=1$),  selection is not needed. ORS and PRS achieve identical performance, which is inferior to that of DT because the relaying in this case is inefficient (power and time resources must be shared between the source and the best relay). However, DT has only one possible signal path, i.e., no diversity,  and its performance curve drops at a rate of $1/\bar{\gamma}^1$. Even though a relay with better CSI is selected from two available relays ($K=2$), the cooperative network can outperform DT thanks to diversity gain. Increasing the network scale with more relays, the superiority of selection, especially PRS, becomes increasingly evident. That is because a diversity order of $K$ is achievable and the performance curve drops at a rate of $1/\bar{\gamma}^K$.

\subsection{Scalability}
Due to the flexibility of a distributed system, the distributed PRS can scale up and down to support dynamic network scale with different numbers of relays. If a new node would participate in a cooperative network, it is first admitted through some mechanisms like admission control and then synchronizes with other nodes. As we can see in Algorithm 1, when RTS/CTS is broadcasted, this node estimates and predicts local CSI, starts a timer, and serves as the best relay if it gets the largest CSI. This procedure is independent and transparent to other relays. Vice versa, when some nodes leave, the remaining relays form a smaller cooperative network that carries out relay selection without any need of modifying the algorithm. Simulation results reveal that the PRS scheme performs well in different numbers of relays, as shown in \figurename \ref{Figure_others}b. 
In the centralized PRS, the relays have no predictor, and only the destination runs a global predictor. The destination knows the network topology and can manage the number of relays employing admission control. Due to the flexibility of deep neural networks, only the dimensions of the input and output layer must be adjusted according to the number of relays. Keeping other hyper-parameters such as two LSTM hidden layers with 25 neurons per layer might raise small derivation on prediction accuracy, but will not cause a breakdown of the cooperative network. Using freshly collected CSI data, an independent DL model is trained off-line to derive optimal hyper-parameters for the changed network, which can be applied to smoothly update the online DL predictor.

\subsection{Computational Complexity}
Last but not least, the complexity of the predictor is quantified to compare with the capacity of COTS computing hardware. As recommended by \figurename \ref{Fig_MSE}, the applied deep neural network has two LSTM hidden layers with $N_h^1=N_h^2=25$. In centralized PRS, the input vector $\textbf{d}_t^{(0)}$ for the global predictor contains $40$ entries due to $K=8$ and $\tau=4$ while the output $\check{\textbf{a}}_d[t+1]$ is a $8$-dimensional vector, corresponding to $N_i=40$ and $N_o=8$. It amounts to $O_{lstm}=25,400$ floating-point operations \emph{per prediction} in terms of (\ref{Eqn_complexity_deepLSTM}). For distributed selection, the local predictor at each relay is much simpler due to a reduced dimension of input and output ($N_i=5$ and $N_o=1$), and therefore we skip it. Since the interval of prediction step is $1\mathrm{ms}$,  the frequency of prediction equals to $f_p=1000$, resulting in \SI{25.4}{\mega FLOPS}. In comparison with off-the-shelf digital signal processors (DSPs), e.g., TI C6678, which provides a computation capacity of up to \SI{179}{\giga FLOPS}, the required computing resource occupies approximately $0.014\%$ of a single DSP chip. Taking into account its back-compatibility to legacy hardware and its applicability to low-cost IoT devices, we further check low-end DSPs. Utilizing TI C6748 that has  computation power of  \SI{2.7}{\giga FLOPS} as an example, the resource required by the predictor is around  $1\%$.   In addition to DSPs, similar results on devices with central processing unit (CPU) or graphical processing unit (GPU) can be expected. In contrast to the ORS system, the proposed scheme does not bring extra overhead on signal transmission. The increase of computational complexity arises only from channel prediction. If increasing $1\%$ load on a running DSP, the increase of energy consumption is marginal. In a nutshell, the complexity of the DL-based channel predictor applied for PRS, as well as its associated energy consumption, is quite affordable, if not negligible.

\section{Conclusions}
In this article, we proposed and analyzed a deep-learning-aided cooperative diversity method for mobile terminals without an antenna array to cultivate the benefit of spatial diversity. A deep recurrent neural network was deliberately built to improve the timeliness of channel state information. The predictor is applicable for any kind of wireless fading statistics, while specific examples for Rayleigh and Rician fading were given. It achieves the optimal performance with full diversity on the order of the number of cooperating relays in slow fading wireless environments, and it substantially outperforms the existing schemes in fast time-varying channels.  It supports both amplify-and-forward and decode-and-forward relaying strategies, and adapts to both distributed and centralized relay selection. Simply inserting a predictor between the channel estimator and relay selector, an ORS system can be transparently upgraded to a PRS system without any other modifications, making it compatible with the existing systems and standards. By selecting a single opportunistic relay, it inherits the simplicity of ORS and avoids multi-relay coordination and synchronization.  The computational complexity and energy consumption arising from fading channel prediction is negligible. Moreover, it is robust enough against additive noise, synchronization error, mobility, and different network scale.
From the perspective of \textit{performance},   \textit{compatibility},\textit{ complexity}, \textit{robustness}, and \textit{scalability}, it is viewed as an excellent candidate for immediate implementation in next-generation cooperative networks.

\appendices
\section{Derivation of Lemma \ref{Lemma_02}}
From \cite{Ref_MGFcapacity}, we know that
\begin{IEEEeqnarray}{lll}
\label{Eqn_log1x}
\log_2 (1+\gamma) = &  - \frac{1}{\ln(2)} \int_0^{\infty} \left[ \frac{\partial}{\partial s} e^{-s \gamma} \right] \times & \\\nonumber
  & H_{3,2}^{1,2} \left[ \frac{1}{s} \left|  \begin{aligned}
         &(1,1),  (1,1),  (1,1) \\
         &(1,1), (0,1)
                          \end{aligned}  \right. \right] ds, &
\end{IEEEeqnarray}
where $H_{3,2}^{1,2}[\cdot]$ is a special mathematical function named Fox's H. It is too complex to solve even with the aid of mathematical software. Another special function called Meijer's G is therefore utilized to replace Fox's H since many software tools such as MATHEMATICA$^{\circledR}$ and MATLAB$^{\circledR}$ have already implemented it. It can be directly invoked to return a numerical value, e.g., $G_{2,1}^{0,2} \left[ \left.  \begin{aligned}
         &1,1\\
         &0
\end{aligned}  \right | 1  \right]{\approx}0.21938$, facilitating the derivation of a closed-form expression. Defining
\begin{equation}\nonumber
\Phi(s) =-H_{3,2}^{1,2} \left[ \frac{1}{s} \left|  \begin{aligned}
         &(1,1),  (1,1),  (1,1) \\
         &(1,1), (0,1)
                          \end{aligned}  \right. \right]=-G_{2,1}^{0,2} \left[ \left.  \begin{aligned}
         &1,1\\
         &0
\end{aligned}  \right | \frac{1}{s}  \right]\end{equation}
and substituting (\ref{Eqn_log1x}) into $C=\int_0^{\infty}\log_2 (1+\gamma)f(\gamma)d\gamma $, yields
\begin{equation}
\label{Eqn_ergocapactiy2}
C=\frac{1}{\ln(2)} \int_0^{\infty} \int_0^{\infty} \left[ \frac{\partial}{\partial s} e^{-s \gamma} \right] \Phi(s)  f(\gamma)d\gamma ds.
\end{equation}
Since $M_{\gamma}(s){=}\int_0^{\infty}e^{-s\gamma}f(\gamma)d\gamma$, (\ref{Eqn_ergocapactiy2}) can be transformed into
\begin{equation}
\label{Eqn_ergocapactiy3}
C=\frac{1}{\ln(2)} \int_0^{\infty} \left[ \frac{\partial M_{\gamma}(s) }{\partial s}  \right] \Phi(s)  ds.
\end{equation}
To avoid the intractability of taking integral, utilizing the Gauss-Chebyshev quadrature shown in \cite{Ref_MGFcapacity}, (\ref{Eqn_ergocapactiy3}) is transformed to (\ref{Eqn_ergocapactiy_approx}).



\bibliographystyle{IEEEtran}
\bibliography{IEEEabrv,Ref_TVT}

\begin{IEEEbiography}
[{\includegraphics[width=1in,height=1.25in,clip,keepaspectratio]{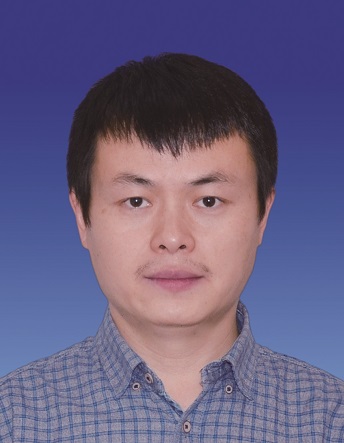}}]
{WEI JIANG} (M'09--SM'19) received the Ph.D. degree in Computer Science from Beijing University of Posts and Telecommunications in 2008.
From 2008 to 2012, he was with the 2012 Laboratory, HUAWEI Technologies. From 2012 to 2015, he was with Institute of Digital Signal Processing, University of Duisburg-Essen, Germany. Since 2015, he is a Senior Researcher with German Research Center for Artificial Intelligence (DFKI), which is the biggest European AI research institution and is the birthplace of ``Industry 4.0" strategy. Meanwhile, he is a Senior Lecturer with University of Kaiserslautern, Germany. He is the author of three book chapters and over 60 conference and journal papers, holds around 30 granted patents, and participated in a number of EU and German research projects. He is an Associate Editor for \textit{IEEE Access} and is a Moderator for IEEE TechRxiv.
\end{IEEEbiography}

\begin{IEEEbiography}
[{\includegraphics[width=1in,height=1.25in,clip,keepaspectratio]{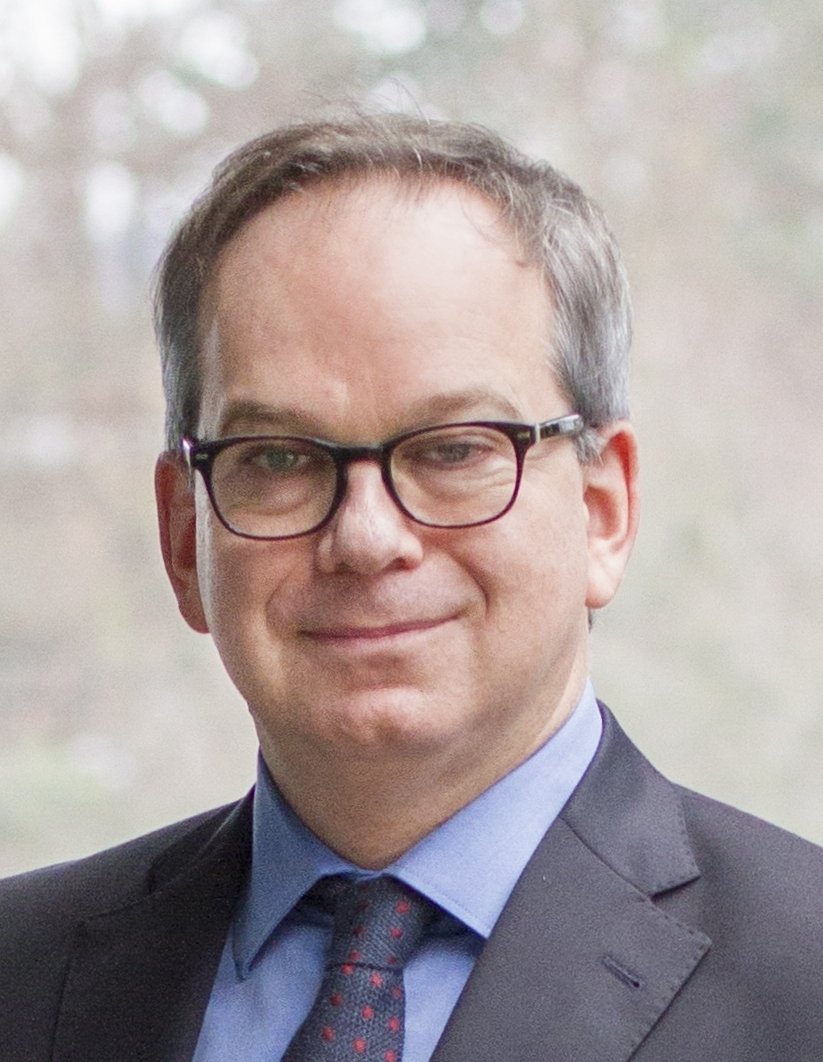}}]
{Hans D. Schotten} (S'93--M'97) received the Ph.D. degrees from the RWTH Aachen University of Technology, Germany, in 1997. From 1999 to 2003, he worked for Ericsson. From 2003 to 2007, he worked for Qualcomm. He became manager of a R\&D group, Research Coordinator for Qualcomm Europe, and Director for Technical Standards. In 2007, he accepted the offer to become the full professor at the University of Kaiserslautern. In 2012, he - in addition - became scientific director of the German Research Center for Artificial Intelligence (DFKI) and head of the department for Intelligent Networks. Professor Schotten served as dean of the department of Electrical Engineering of the University of Kaiserslautern from 2013 until 2017. Since 2018, he is chairman of the German Society for Information Technology and member of the Supervisory Board of the VDE. He is the author of more than 200 papers and participated in 30+ European and national collaborative research projects.
\end{IEEEbiography}
\end{document}